\documentclass[aps,prb,twocolumn,superscriptaddress,longbibliography]{revtex4-2}

\usepackage{graphicx,bm,amssymb,amsmath,xcolor,pifont,soul,multirow}
\usepackage[linktocpage=true,colorlinks=true,pdfborder={0 0 0},linkcolor=blue,citecolor=blue,filecolor=yellow,urlcolor=blue,bookmarks,pdfauthor={},]{hyperref}
\usepackage[normalem]{ulem}
\usepackage{epstopdf}
\usepackage{epsfig}
\usepackage{blkarray}

\usepackage{array,booktabs,longtable,tabularx}
\newcolumntype{L}{>{\raggedright\arraybackslash}X}
\usepackage{ltablex}
\usepackage{siunitx}
\usepackage{titlesec}
\usepackage{lipsum}

\def\bk{{\bf k}}

\def\bq{{\bf q}}

\def\>{\rangle}
\def\<{\langle}
\usepackage{amsmath}

\setstcolor{red}

\titleformat{\subsubsection}
  {\normalfont}{\thesubsubsection}{1em.}{}
\renewcommand{\thesubsubsection}{\roman{subsubsection}}

\begin{document}
\title{First-principles design of ambient-pressure Mg$_x$B$_2$C$_2$ and Na$_x$BC superconductors}

\author{Charlsey R. Tomassetti}
\affiliation{Department of Physics, Applied Physics, and Astronomy, Binghamton University-SUNY, Binghamton, New York 13902, USA}
\author{Daviti Gochitashvili}
\affiliation{Department of Physics, Applied Physics, and Astronomy, Binghamton University-SUNY, Binghamton, New York 13902, USA}
\author{Christopher Renskers}
\affiliation{Department of Physics, Applied Physics, and Astronomy, Binghamton University-SUNY, Binghamton, New York 13902, USA}
\author{Elena R. Margine}
\email{rmargine@binghamton.edu}
\affiliation{Department of Physics, Applied Physics, and Astronomy, Binghamton University-SUNY, Binghamton, New York 13902, USA}
\author{Aleksey N. Kolmogorov}
\email{kolmogorov@binghamton.edu}
\affiliation{Department of Physics, Applied Physics, and Astronomy, Binghamton University-SUNY, Binghamton, New York 13902, USA}
\date{\today}

\begin{abstract}
We employ {\it ab initio} modeling to investigate the possibility of attaining high-temperature conventional superconductivity in ambient-pressure materials based on the known MgB$_2$C$_2$ and recently proposed thermodynamically stable NaBC ternary compounds. The constructed $(T,P_{\textrm M})$ phase diagrams (M = Mg or Na) indicate that these layered metal borocarbides can be hole-doped via thermal deintercalation that has been successfully used in previous experiments to produce Li$_{1>x\gtrsim0.5}$BC samples. The relatively low temperature threshold required to trigger NaBC desodiation may help prevent the formation of defects shown recently to be detrimental to the electron-phonon coupling in the delithiated LiBC analog. According to our numerical solutions of the anisotropic full-bandwidth Migdal-Eliashberg equations, the proposed Mg$_x$B$_2$C$_2$ and Na$_x$BC materials exhibit superconducting critical temperatures between 43~K and 84~K. At the same time, we demonstrate that buckling of defect-free honeycomb BC layers, favored in heavily-doped Na$_x$BC compounds, can substantially reduce or effectively suppress the materials’ potential for MgB$_2$-type superconductivity.

\end{abstract}	

\maketitle

\section{Introduction}
\label{sec:introduction}

The naturally hole-doped stoichiometric MgB$_2$ with honeycomb boron layers has served as a blueprint for designing ambient-pressure high-$T_{\rm c}$ superconductors ever since the unexpected 2001 discovery of the material’s conventional superconductivity at 39 K~\cite{Nagamatsu2001}. It was quickly established that no other known diborides of Al or transition metals display the quasi-2D electronic features responsible for the strong electron-phonon (e-ph) coupling, while the possibility of adding new members to the large MB$_2$ family is prohibited by unfavorable thermodynamics at ambient conditions (M = Li, Cu, Ag, Au, {\it etc.})~\cite{Pelleg2007, ak09}. Expansion of the search space to other metal-boron compositions ({\it e.g.}, MB) or ternary metal borides ({\it e.g.}, Mg$_x$M$_{1-x}$B$_2$ and Li$_x$M$_y$B) further demonstrated the difficulty of obtaining stable materials with the desired hole-doped covalent bonds~\cite{ak09,ak08,ak10,ak14,ak30, Cava2003,  Bianconi2007, Karpinski2008, Parisiades2009}. Meanwhile, graphite intercalation compounds (GICs) with fully filled $\sigma$ states display a different e-ph coupling mechanism and generally lower $T_{\rm c}$~\cite{Calandra2005,Mazin2005,Margine2016}.

The electron-deficient honeycomb frameworks with alternating B and C atoms possess the signature quasi-2D electronic states and hard vibrational modes to be MgB$_2$-type superconductors. Since the only two metal borocarbides with this morphology, LiBC~\cite{worle1995libc} and MgB$_2$C$_2$~\cite{worle1994}, are semiconductors in the stoichiometric form, several studies have been dedicated to investigating their hole-doped derivatives. Li$_{1>x}$BC had been predicted in early computational works to superconduct at temperatures as high as 100~K~\cite{Ravindran2001, Rosner2002, Dewhurst2003}, but successful experimental efforts to delithiate Li$_x$BC unfortunately resulted in no detectable $T_{\rm c}$~\cite{Bharathi2002,  Nakamori2003, Zhao2003,Fogg2003a,Fogg2003b,Fogg2006, Kalkan2019}. The lack of superconductivity has been attributed to the likely presence of defects in the BC layers that become thermodynamically favored at low Li concentrations~\cite{Fogg2006, Tomassetti2024}. Reintercalation of Li$_x$BC with other group I or II elements has been  identified in our recent work as a possible route of obtaining elusive MgB$_2$ analogs~\cite{Tomassetti2024}. 

Computational studies examining hole-doped MgB$_2$C$_2$, through removal and/or replacement of Mg with alkali metals~\cite{Verma2003, Pham2023, Spano2005, Mori2002}, also noted the materials’ potential for high-$T_{\rm c}$ superconductivity. In particular, Span\`o {\it et al.}~\cite{Spano2005} considered various (Mg,Li)B$_2$C$_2$ and (Mg,Na)B$_2$C$_2$ phases and concluded that a substitution of Mg by Li is energetically favorable in hole-doped MgB$_2$C$_2$. The most extensive experimental investigation of the compound’s derivatives was performed by Mori and Takayama-Muromachi in 2004~\cite{Mori2004}. The appearance of Pauli magnetism in the synthesized Mg$_{0.5}$Li$_{0.8}$B$_2$C$_2$ material indicated that the sample became metallic, but no superconductivity was observed down to 1.8~K. Attempts at high pressure synthesis of (Mg,Li)B$_2$C$_2$ and boron doping of the B/C nets in MgB$_2$C$_2$ were unsuccessful, resulting in large amounts of neighboring phases, LiBC and MgB$_2$~\cite{Mori2004}. To the best of our knowledge, there have been no reported attempts to deintercalate MgB$_2$C$_2$ through high-temperature annealing as has been done with LiBC~\cite{Fogg2003a, Zhao2003, Fogg2006, Kalkan2019}.

\begin{figure*}[t!]
   \centering
\includegraphics[width=0.95\textwidth]{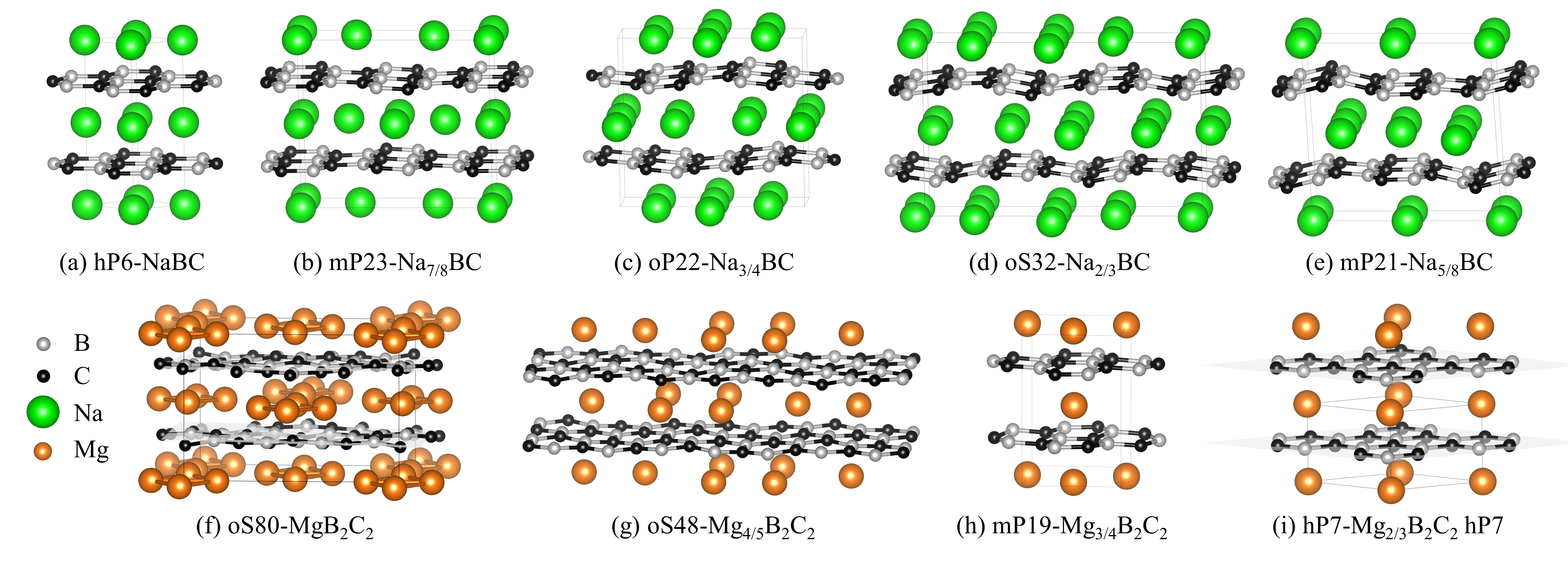}
    \caption{\label{fig-01} Crystal structures of select examined layered Na$_x$BC and Mg$_x$B$_2$C$_2$ phases denoted with Pearson symbols. The fully intercalated compounds are (a) NaBC with perfectly flat BC layers previously proposed to be a low-temperature ground state~\cite{Tomassetti2024} and (f) MgB$_2$C$_2$ with BC layers buckled around rhombus-shaped Mg patches first observed in 1994~\cite{worle1994}.}
\end{figure*}

The recently uncovered thermodynamic stability of NaBC at low temperatures~\cite{Tomassetti2024}, in a simple hexagonal structure~\cite{Miao2016}, offers another pathway for designing high-$T_{\rm c}$ metal borocarbides. Prior theoretical work simulated holes in NaBC through desodiation~\cite{Singh2002} or substitution of C by B~\cite{Miao2016}, and predicted the hole-doped NaBC to have high critical temperatures in both cases ({\it e.g.}, 35~K for NaB$_{1.1}$C$_{0.9}$ using the Allen-Dynes modified McMillian formula and $\mu^*$=0.1~\cite{Miao2016}). The discovery of a new NaB$_5$C compound, as recently as 2021~\cite{Delacroix2021}, brings up questions about what other materials may exist in this ternary system and what properties they may display.

In this work, we use density functional theory (DFT) to analyze thermodynamic stability and superconducting properties of the Na$_{1>x}$BC and Mg$_{1>x}$B$_2$C$_2$ compositional subspaces. We identify the most favorable layered configurations, map out the ($T$,$P_{\rm M}$) synthesis conditions needed to thermally deintercalate the ternary compounds, and calculate the superconducting $T_{\rm c}$ of the hole-doped materials using the anisotropic Midgal-Eliashberg (aME) formalism. Our findings indicate that NaBC and MgB$_2$C$_2$, shown in Fig.~\ref{fig-01}, are promising precursors for ambient-pressure synthesis of high-$T_{\rm c}$ conventional superconductors.

\section{Methods}
\label{sec:methods}
{\small VASP}~\cite{Kresse1996} was used to conduct the stability analysis of Na-B-C and Mg-B-C phases using projector augmented wave potentials~\cite{Blochl1994} and a 500~eV plane-wave cutoff. Due to the known importance of dispersive interactions in layered materials~\cite{ak06,Lebegue2010,ak30,rvv-1}, we used the optB86b-vdW functional~\cite{optB86b}, and checked the sensitivity of the results to the DFT approximations with the optB88-vdW~\cite{optB88} and r2SCAN+rVV10~\cite{r2scan, rvv-1} functionals. All structures were evaluated with dense ($\Delta k \sim 2 \pi \times 0.025$~\AA$^{-1}$)  Monkhorst-Pack $\bk$-meshes~\cite{Monkhorst1976}.

Global structure searches were performed with an evolutionary algorithm implemented in the MAISE package~\cite{maise}. In fixed-composition runs, randomly initialized 16-member populations with up to 22 atoms per unit cell were evolved for up to 250 generations using standard mutation and crossover operations~\cite{maise}. The thermodynamic corrections due to vibrational entropy were calculated within the finite displacement methods implemented in PHONOPY~\cite{Togo2015}. We employed supercells between 56 and 168 atoms, applying 0.1~\AA\ displacements within the harmonic approximation. Previous quasi-harmonic approximation results for related Li-B-C, Li-Na-B-C, and Na-Sn materials~\cite{Kharabadze2023, thorn2023, Tomassetti2024} show that volume expansion has a negligible effect on the formation free energies in the considered temperature range. A combinatorial screening method was used to sequentially remove intercalants and leave only non-equivalent configurations.  Equivalence was evaluated with our structural fingerprint based on the radial distribution function~\cite{ak16,maise}.

The \textsc{Quantum} {\small ESPRESSO} package~\cite{QE} was used for calculating properties related to superconductivity. We employed the optB86b-vdW and optB88-vdW functionals~\cite{optB86b, optB88, Thonhauser2007, Thonhauser2015, Berland2015, Langreth2009} and norm-conserving pseudopotentials from the Pseudo Dojo library~\cite{Dojo2018} generated with the relativistic PBE parametrization~\cite{PBE}. A plane-wave cutoff value of 100~Ry, a Methfessel-Paxton smearing~\cite{Methfessel1989} value of 0.02~Ry, and $\Gamma$-centered Monkhorst-Pack~\cite{Monkhorst1976} \textbf{k}-meshes were used to describe the electronic structure. The lattice parameters and atomic positions were relaxed until the total energy was converged within $10^{-6}$~Ry and the maximum force on each atom was less than $10^{-4}$ Ry/\AA. The dynamical matrices and the linear variation of the self-consistent potential were calculated within density-functional perturbation theory~\cite{Baroni2001} on irreducible sets of regular \textbf{q}-meshes. The optimized lattice parameters for the investigated phases and the \textbf{k}- and \textbf{q}-meshes used are reported in Table~S1~\cite{SM}. 

\begin{figure*}[t!]
   \centering
\includegraphics[width=0.90\textwidth]{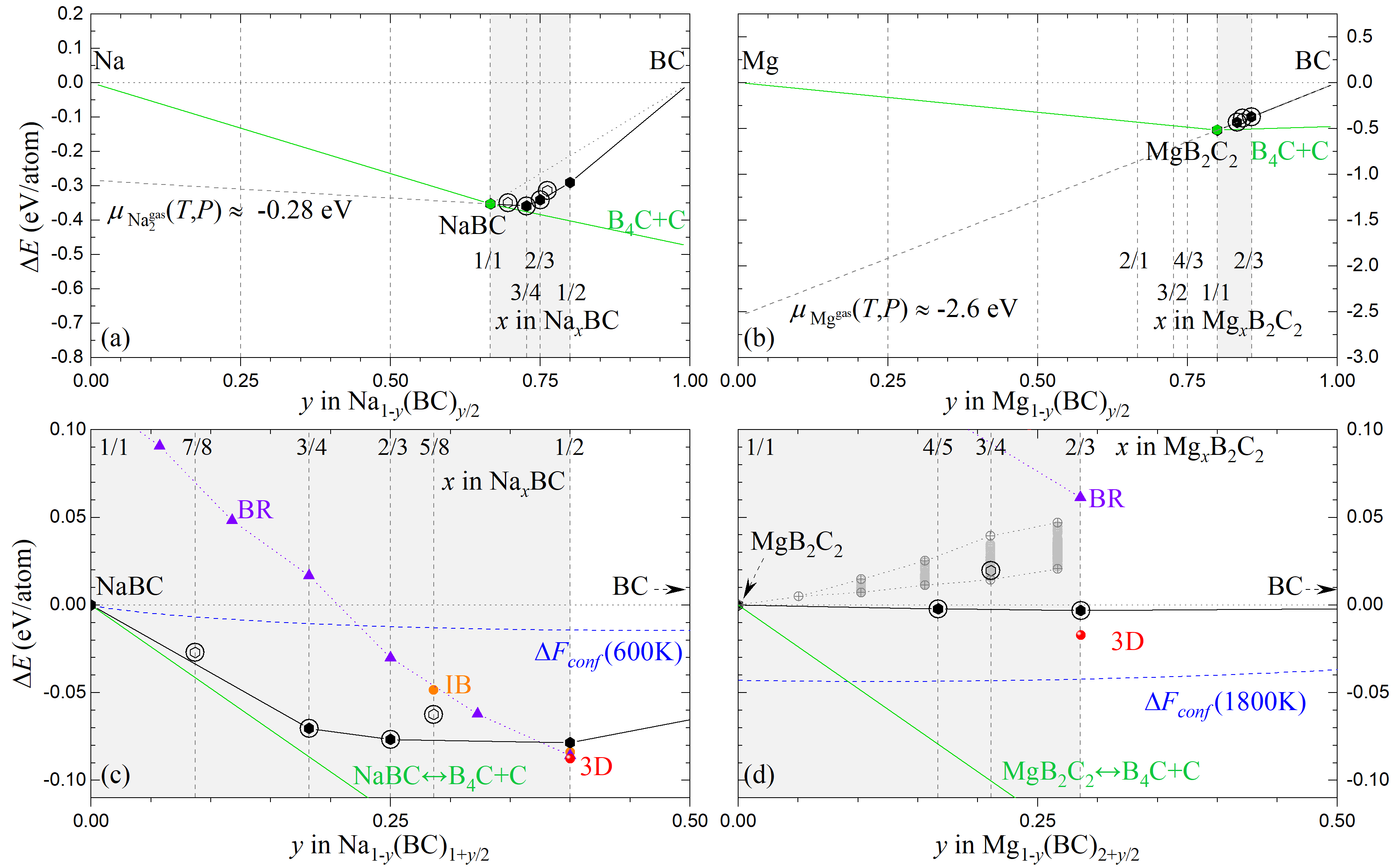}
    \caption{\label{fig-02} Relative energies of Na$_x$BC and Mg$_x$B$_2$C$_2$ phases referenced to a hypothetical fully deintercalated BC phase and (a),(b) the elemental ground states or (c),(d) the fully intercalated ternary ground states. The solid green lines define the global convex hulls, the solid black lines mark the boundaries of the local convex hulls for deintercalated honeycomb derivatives, the dashed blue lines show the estimated configurational entropy contributions from metal disorder, and the dashed gray lines point to the chemical potentials of Na$_2^{\rm gas}$ (Mg$^{\rm gas}$) needed to destabilize the starting NaBC (MgB$_2$C$_2$) phase. The solid symbols denote metal borocarbides with honeycomb (black hexagons), bond rotation (BR, purple triangles), interlayer bridging (IB, orange circles), and fully connected (3D, red circles) morphologies of the BC covalent networks. The crossed gray circles in panel (d) correspond to deintercalated phases derived from the large experimental MgB$_2$C$_2$ structure. The circle points highlight materials examined for their superconducting properties. The shaded areas correspond to compositions that should be accessible via deintercalation.  }
\end{figure*}

The e-ph interactions and superconducting properties were evaluated with the EPW code~\cite{Giustino2007, EPW2016, Margine2013, EPW2023}. The Wannier interpolation~\cite{WANN1, WANN2, WANN3} was performed on uniform $\Gamma$-centered \textbf{k}-grids (see Table~S1~\cite{SM}) with the Wannier90 code~\cite{WANN1, WANN2, WANN3} in library mode. We considered $2p$ orbitals for every C or B atom as projections for the maximally localized Wannier functions to accurately describe the electronic structure of all the compounds under investigation. The anisotropic full-bandwidth~\cite{EPW2023,Lucrezi2024} equations were solved with a sparse intermediate representation of the Matsubara frequencies~\cite{mori2024} on fine uniform $\bk$- and $\bq$-point grids (see Table~S1~\cite{SM}), with an energy window of $\pm 0.2$~eV around the Fermi level. The Coulomb $\mu^*$ parameter was chosen to be 0.20, which ensures good agreement between measured and computed aME $T_{\rm c}$ values for MgB$_2$~\cite{Kafle2022}. Visualizations of crystal structures and Fermi surfaces were created with VESTA~\cite{vesta} and FermiSurfer~\cite{fermisurfer}, respectively. To resolve different phases at the same composition, we use Pearson symbols and space groups when needed. Full structural information for relevant DFT-optimized Na$_x$BC or Mg$_x$B$_2$C$_2$ phases is provided as CIF files in the Supplementary Material.


\section{Results and discussion}

\label{sec:Stability}
\subsection{Stability}

Complementary structure search strategies shown to be effective in the exploration of related metal borocarbides~\cite{Tomassetti2024} were employed to examine the Na$_x$BC and Mg$_x$B$_2$C$_2$ subspaces with the ultimate aim of determining the composition ranges where the materials should remain layered and relatively defect-free after hole-doping. First, we conducted a combinatorial screening by generating different NaBC and MgB$_2$C$_2$ (super)cells and systematically removing metal ions, leaving only the non-equivalent configurations. The supercells included standard expansions of the fully-filled hP6 structure, known to be the ground state for LiBC and NaBC, such as $\sqrt{3}\times\sqrt{3}\times1$, $2\times2\times1$, and orthorhombic supercells of increasing size up to 54 atoms with both AA and AA’ stackings of the BC sheets. We also proceeded with systematic deintercalation of the oS80 ground state of MgB$_2$C$_2$ down to $x=11/16$. The full set of considered layered phases contained over 700 distinct metal decorations.

The formation energies of the best candidate hole-doped structures (Fig.~\ref{fig-02}(a)-(b)) fall above the global convex hull as defined by the respective combinations of NaBC (MgB$_2$C$_2$), B$_4$C, and C materials. However, the successful deintercalation experiments on LiBC have established that the BC framework retains its layered morphology at high temperatures (up to 1770 K) and low metal concentrations (down to at least $x=0.5$)~\cite{Fogg2006}. This indicates that the high kinetic barriers associated with rebonding of the covalent layers are sufficiently high to keep the materials from decomposing into other products under a wide range of experimental conditions. Hence, we examine the feasibility of the intercalated M$_x$BC phases within the kinetically restricted subspace and, for convenience, display their relative stability with respect to NaBC (MgB$_2$C$_2$) and a fully deintercalated hypothetical layered BC, as was done in our previous study on Li$_x$BC~\cite{ Kharabadze2023}. Fig.~\ref{fig-02}(c) and (d) shows that the best Mg$_x$B$_2$C$_2$ phases, just as in the LiBC case, stay close to the MgB$_2$C$_2$ $\leftrightarrow$ BC tie-line, but the Na$_x$BC derivatives fall way below the corresponding reference line and are actually within 25 meV/atom, for $x$ down to 3/4, of being globally stable relative to NaBC, B$_4$C, and C. For reference, we show configurational entropy contributions estimated as $\Delta F_{\textrm{conf}}=kT[x\ln(x)+(1-x)\ln(1-x)]/(2+x)$ as functions of $y=(1-x)/(1+x/2)$ in Fig.~\ref{fig-02}(c) or $y=(1-x)/(2+x/2)$ in Fig.~\ref{fig-02}(d), at the expected deintercalation temperatures.

To analyze the preferred Mg arrangement after deintercalation, we first considered the conventional unit cell of the oS80-MgB$_2$C$_2$ ground state, where the Mg atoms are grouped in rhombus-shaped patches with two non-equivalent sites at the acute- and obtuse-angled corners (Fig.~\ref{fig-01}(f)). The relative energies of Mg-depleted structures are contained in a well-defined wedge (Fig.~\ref{fig-02}(d)), with the more stable configurations featuring Mg atoms in the obtuse-angle sites. Structures derived from hP6-Mg$_2$B$_2$C$_2$ supercells with more even distributions of Mg atoms prove to be more favorable for compositions below $\sim$ 0.95, and one can expect the material to quickly depart from the complex rhombus-patterned arrangement of Mg atoms upon deintercalation. 

\begin{figure}[!t]
   \centering
\includegraphics[width=0.48\textwidth]{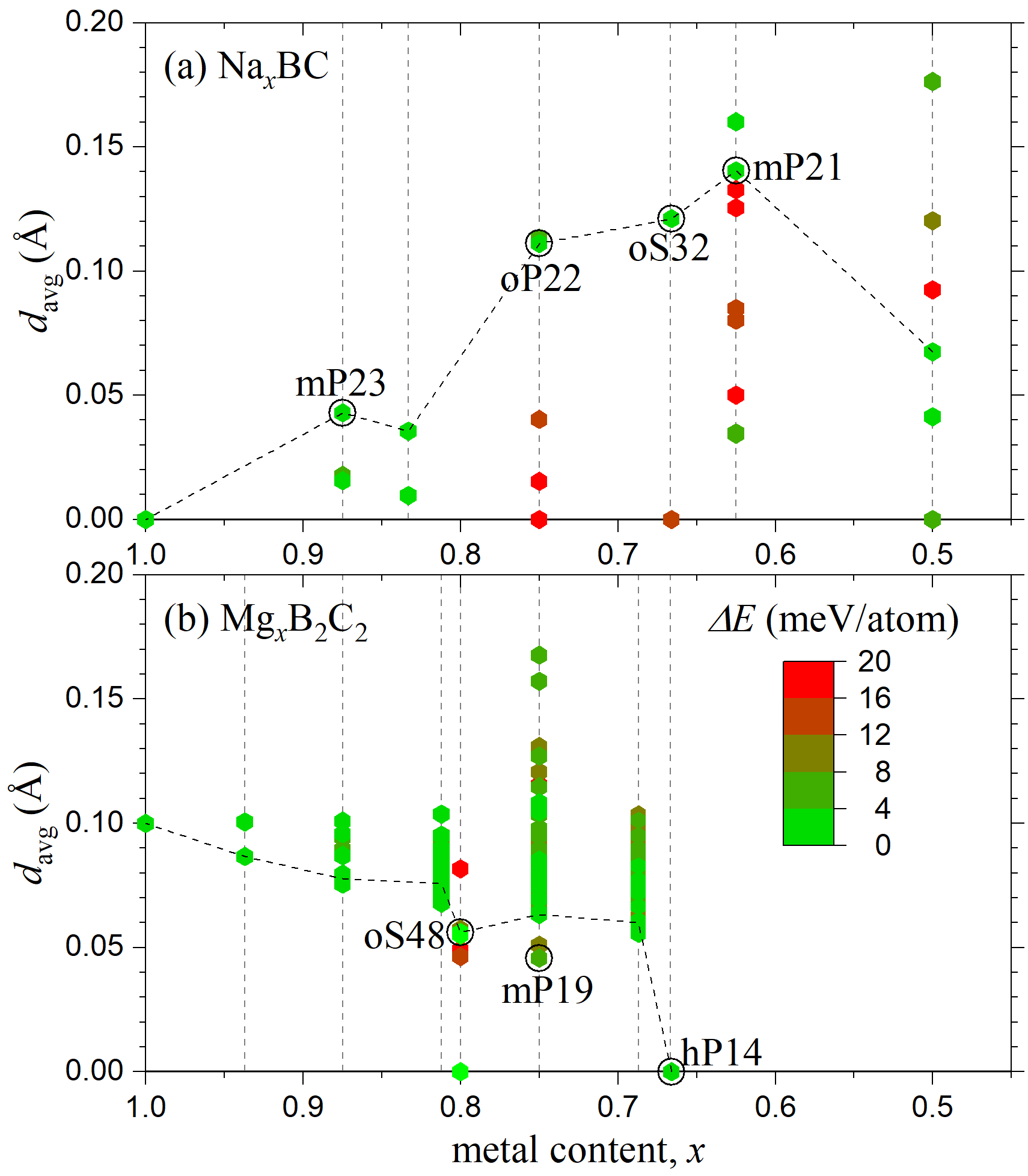}
    \caption{\label{fig03} Average distortion of BC layers ($d_{\rm avg})$ at different compositions $x$ in (a) Na$_x$BC and (b) Mg$_x$B$_2$C$_2$. The color map shows the energy of the considered phases within a 20 meV/atom window relative to the most stable structure at each composition, which are connected with dashed black lines. The circled phases were examined for superconducting properties with aME calculations.}
\end{figure}

The amount and distribution of the intercalants can influence the compound’s superconducting properties by defining not only the doping level of the BC sheets but also the degree of their buckling. The maximum B or C out-of-plane displacements in the synthesized MgB$_2$C$_2$ material~\cite{worle1994} reaches $d_{\rm {max}}$ = 0.20~\AA. To illustrate the dependence of the corrugation on the metal type, concentration, and distribution, we collected the DFT data for all constructed Na$_x$BC and Mg$_x$B$_2$C$_2$ phases in Fig.~\ref{fig03}. Focusing on the most stable structures in each ternary, we observe different trends for covalent framework buckling upon extraction of the larger alkali and smaller alkaline-earth metals. The BC layers start as perfectly flat in the fully loaded NaBC, become progressively more distorted with decreasing Na content reaching $d_{\rm {avg}}=0.14$~\AA\ at $x=5/8$, and finally settle with an intermediate corrugation value of 0.07~\AA\ at $x=1/2$. The MgB$_2$C$_2$ material, being half-filled, possesses a significantly corrugated BC network with $d_{\rm {avg}}=0.10$~\AA, exhibits numerous nearly degenerate configurations with gradually decreasing buckling down to 0.06~\AA\ at $x=11/16$, and has perfectly flat layers in a high-symmetry crystal structure with a uniform distribution of Mg ions at $x=2/3$. Some configurations in both Na and Mg sets led to the natural formation of interlayer C-C bonds upon local relaxation but none of these phases were found to be preferred in the composition range considered. 

Additionally, we constructed layered structures with both AA and AA' stackings incorporating rotated B-C bonds. This defect type was theorized to occur in Li$_x$BC~\cite{Fogg2006, Tomassetti2024} and could be responsible for the absence of superconductivity in the synthesized samples. The most stable stoichiometric ternary phases with rotated bonds, mP15-MgB$_2$C$_2$ and mP36-NaBC, are well above the corresponding ground states, by 151 meV/atom and 119 meV/atom, respectively. In Mg$_x$B$_2$C$_2$, the relative energy of the defective structures with respect to MgB$_2$C$_2$ and BC decreases almost linearly with metal content $x$ but remains well above the energy of the best ordered honeycomb phases, {\it e.g.}, by 61 meV/atom, at the lowest $x=2/3$ end (see Fig.~\ref{fig-02}(d)). In Na$_x$BC, a similar linear trend as a function of composition makes the structures with rotated bonds more competitive, with the best one at $x=1/2$ actually becoming slightly more favored, by 7 meV/atom, over the ordered honeycomb counterparts.

Previous computational work on the Li$_x$BC system also showed that non-layered structures with bridged or fully-connected 3D frameworks can become favored once the metal concentration falls below a certain level ($x \sim 2/3$)~\cite{Kharabadze2023, Tomassetti2024}, but experiments indicate that Li$_x$BC remains layered down to approximately $x\sim 0.45$, at which point the material begins to decompose~\cite{Fogg2006}. To probe the response of the Na$_x$BC and Mg$_x$B$_2$C$_2$ compounds to alternative non-layered configurations, we performed unconstrained evolutionary searches. We found that non-layered morphologies become competitive around $x=1/2$ for Na$_x$BC and $x=2/3$ for Mg$_x$B$_2$C$_2$, see Fig.~\ref{fig-02}(c)-(d).

\begin{figure}[t!]
   \centering
\includegraphics[width=0.48\textwidth]{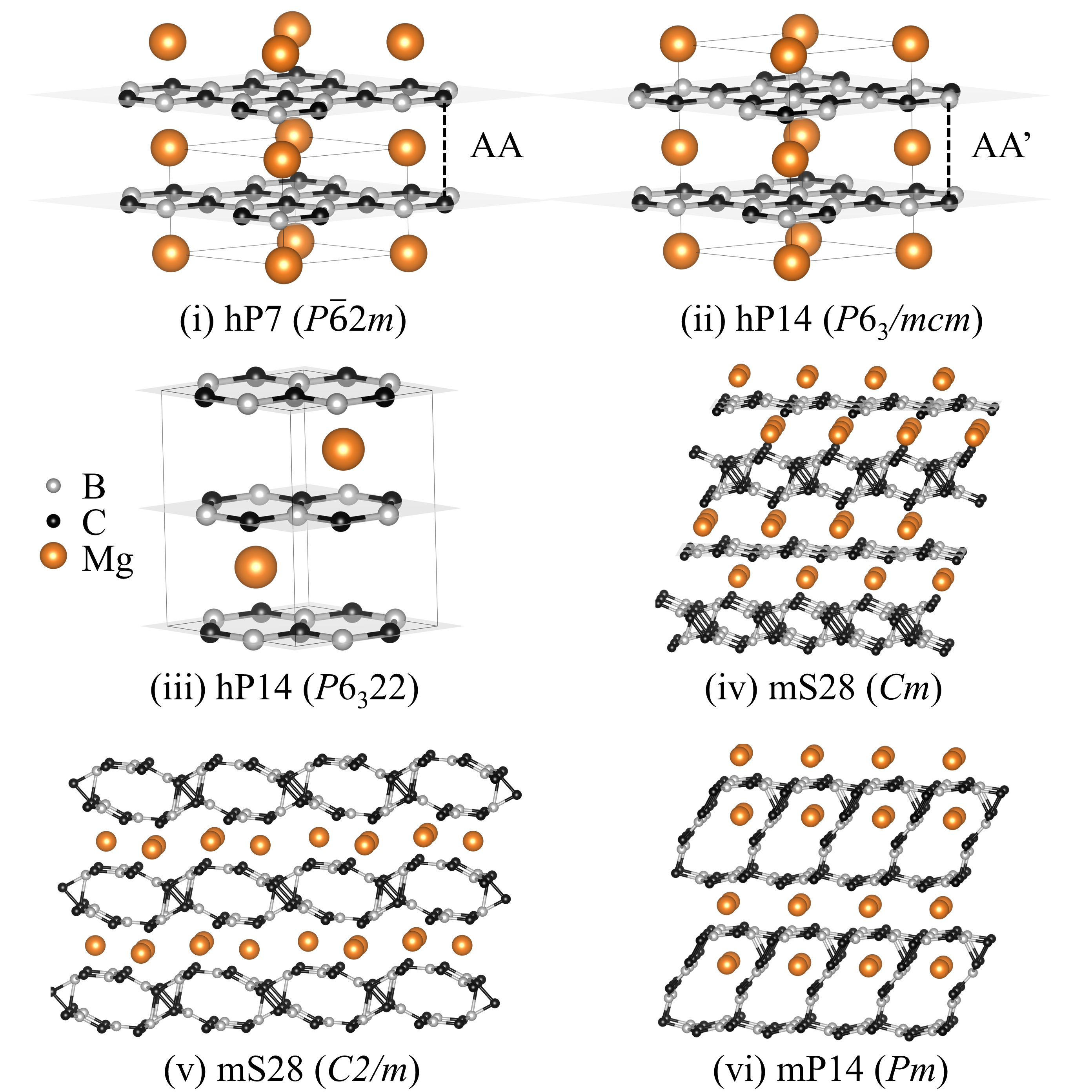}
    \caption{\label{fig-04} Crystal structures of competing Mg$_{2/3}$B$_2$C$_2$ phases proposed in (i) previous work~\cite{Pham2023} and (ii),(vi) the present study.}
\end{figure}

\begin{figure}[t!]
   \centering
\includegraphics[width=0.48\textwidth]{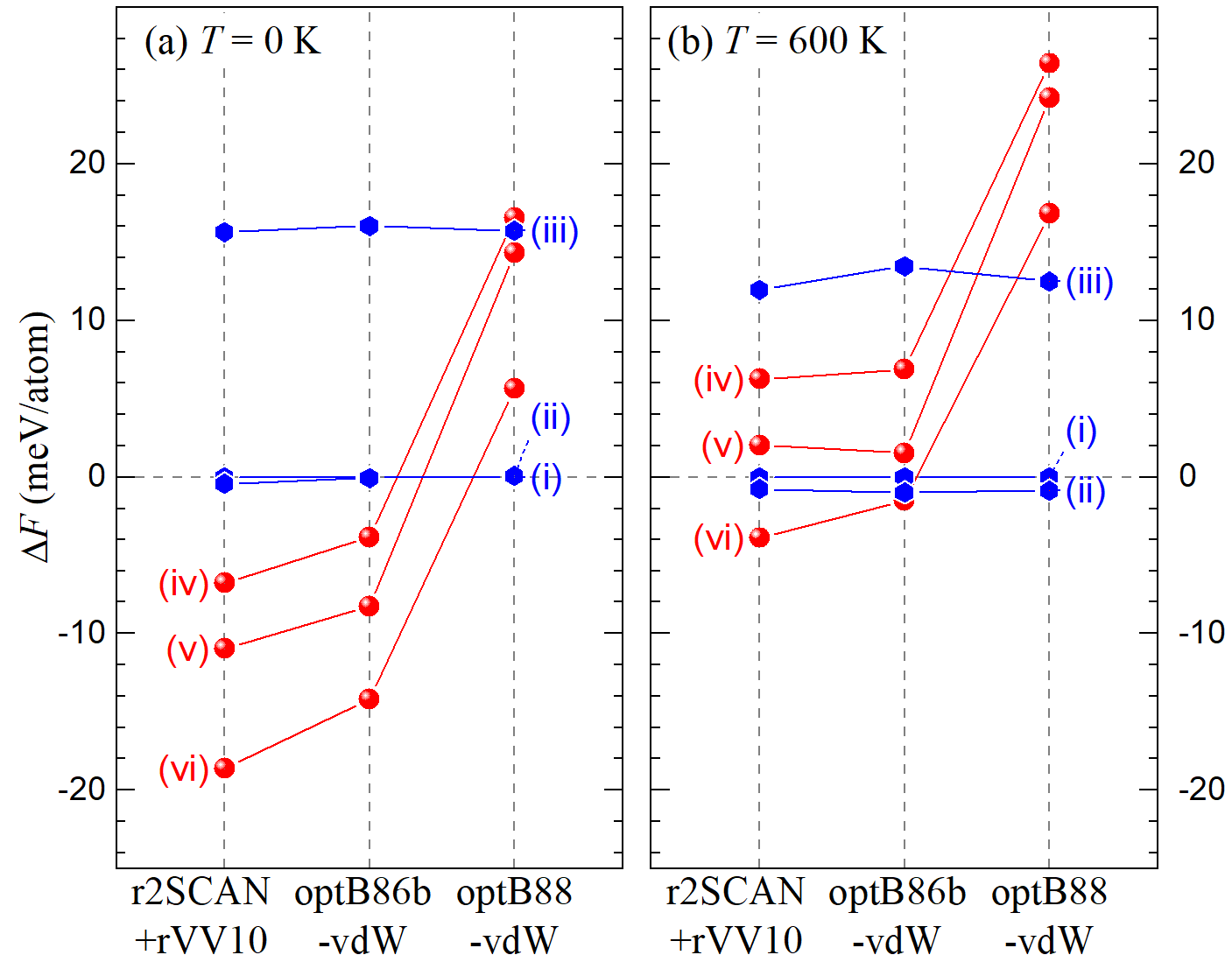}
    \caption{\label{fig-05} Relative free energies of select Mg$_{2/3}$B$_2$C$_2$ polymorphs calculated at (a) $T=0$~K and (b) $T=600$~K with the optB86b-vdW~\cite{optB86b}, optB88-vdW~\cite{optB88}, and r2SCAN+rVV10~\cite{r2scan, rvv-1} functionals. The labeled points correspond to the crystal structures shown in Fig.~\ref{fig-04}. The blue hexagons denote phases with honeycomb BC layers, while the red spheres denote phases with complex re-ordered 3D networks. The latter set is more favored in both optB86b-vdW and r2SCAN+rVV10 approximations up to 600 K.}
\end{figure}

The critical Mg$_{2/3}$B$_2$C$_2$ stoichiometry examined by Pham {\it et al.}~\cite{Pham2023} deserves a closer look. This previous study identified an hP7 structure phase with AA stacking (Fig.~\ref{fig-04}(i)) as the 1:3:3 ground state with the optB88-vdW functional. Our optB88-vdW calculations show that an hP14 structure ($P6_3/mcm$) with the AA' stacking (Fig.~\ref{fig-04}(ii)) is degenerate in energy at 0 K but is slightly preferred by 0.5 meV/atom with the inclusion of the zero-point energy (ZPE) and becomes more stable by 0.9 meV/atom at 600 K with the inclusion of the vibrational entropy. The finding is not unexpected given that the parent MgB$_2$C$_2$ material has the same stacking sequence and that LiBC, with a 5\% shorter interlayer spacing, has a more pronounced preference for AA' over AA by 17 meV/atom at 0 K. To assess the reliability of the observations for these and other relevant polymorphs found with the evolutionary algorithm, we also calculated the free energies with the optB86-vdW and r2SCAN+rVV10 functionals. Fig.~\ref{fig-05} summarizes the structure stability results relative to hP7 obtained with the three exchange-correlation functionals at 0~K and 600~K. The r2SCAN+rVV10~\cite{r2scan, rvv-1} and optB86b~\cite{optB86b} results indicate that several non-layered low-symmetry phases (Fig.~\ref{fig-04}(iv)-(vi)) are noticeably more stable than hP7 or hP14 ($P6_3/mcm$) at 0~K, with mP14 (Fig.~\ref{fig-04}(vi)) remaining preferred at 600~K. The optB88 functional~\cite{optB88}, on the other hand, favors the layered polymorphs. Overall, the three approximations agree that the ordered honeycomb phases remain competitive at this composition, especially at high temperatures, and that the most stable decoration, by at least 15 meV/atom, has the $\sqrt{3}\times\sqrt{3}$ in-plane expansion of the BC primitive unit cell. Therefore, one can indeed expect hP14-Mg$_{2/3}$B$_2$C$_2$ ($P6_3/mcm$) to be the product of MgB$_2$C$_2$ deintercalation at high temperatures.

In the case of Li$_x$BC, deintercalation is driven by the large gain in entropy attained when Li enters the diatomic gas state. As has been done previously~\cite{Kharabadze2023}, we can evaluate the entropy of Na and Mg within the ideal gas model by accounting for either the vibrational, rotational, and translational degrees of freedom for Na$_2^{\textrm{gas}}$, or just the translational degrees of freedom for monatomic Mg$^{\textrm{gas}}$~\cite{chem-potential-textbook}. In Fig.~\ref{fig-02}(a)-(b), we illustrate how NaBC and MgB$_2$C$_2$ should become destabilized when the chemical potentials for Na$_2^{\textrm{gas}}$ and Mg$^{\textrm{gas}}$, relative to their respective bulk ground states, fall below certain free energy values, approximated via extrapolations of the NaBC$\leftrightarrow$Na$_{3/4}$BC or MgB$_2$C$_2$$\leftrightarrow$Mg$_{4/5}$B$_2$C$_2$ lines. The necessary change in the chemical potential for Na$_2^{\textrm{gas}}$ is much smaller in magnitude than those for Li$_2^{\textrm{gas}}$ and Mg$^{\textrm{gas}}$, which indicates that NaBC deintercalation should be easier to achieve.

\begin{figure}[t!]
   \centering
\includegraphics[width=0.40\textwidth]{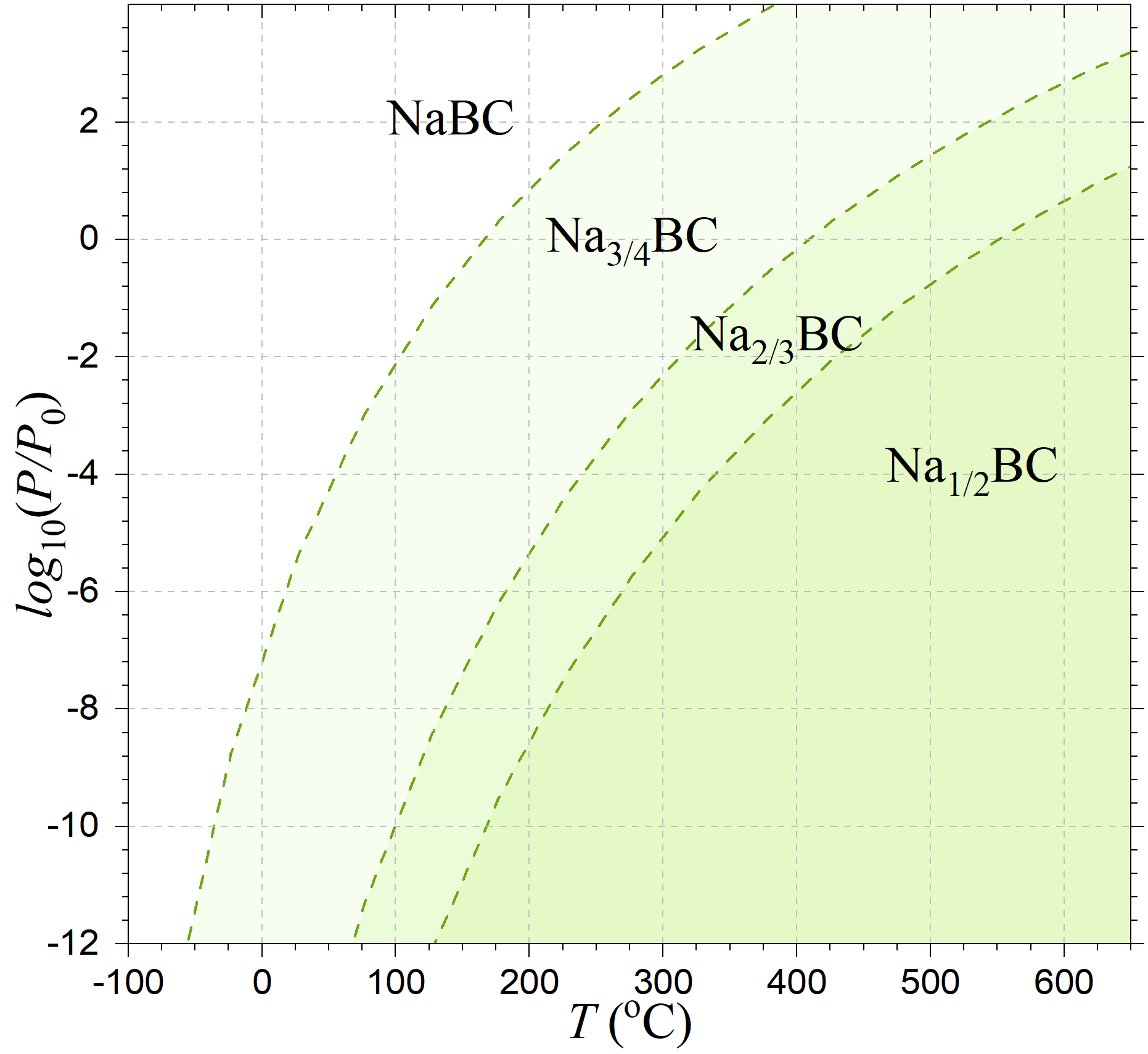}
    \caption{\label{fig-06} Calculated map of phase stability ($T,P_{\rm Na}$) for deintercalated derivatives of NaBC. The phase boundaries correspond to thermodynamic equilibria between neighboring layered Na$_x$BC materials and Na$_2^{\rm gas}$.}
\end{figure}

\begin{figure}[t!]
   \centering
\includegraphics[width=0.40\textwidth]{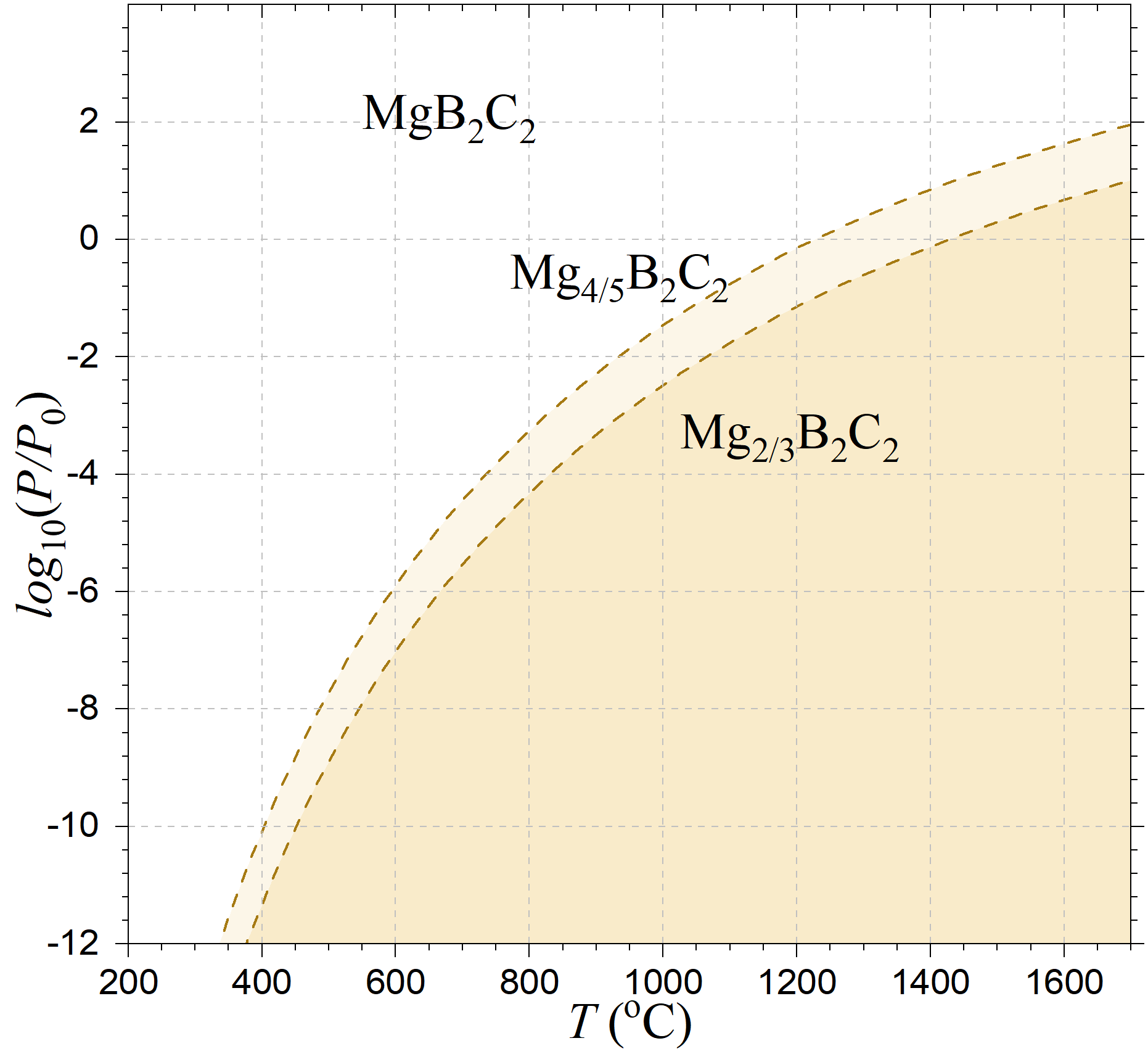}
    \caption{\label{fig-07} Calculated map of phase stability ($T,P_{\rm Mg}$) for deintercalated layered derivatives of MgB$_2$C$_2$. The phase boundaries correspond to thermodynamic equilibria between neighboring layered Mg$_x$B$_2$C$_2$ materials and Mg$^{\rm gas}$.}
\end{figure}

In Figs.~\ref{fig-06} and \ref{fig-07}, we propose $(T,P_{\rm M})$ diagrams demonstrating the required Na$_2^{\textrm{gas}}$ or  Mg$^{\textrm{gas}}$ vapor pressures and temperatures needed to transition to metastable deintercalated products Na$_x$BC and Mg$_x$B$_2$C$_2$. We find the positions of phase boundaries in Na${_x}$BC by analyzing the identified ordered oP22, oS32, and oS80 structures at $x=3/4$, 2/3, and 1/2, respectively. After incorporating the vibrational entropy for the bulk phases, we determined the equilibrium ($T$,$P_{\rm M}$) conditions within the temperature range of 200-2000 K by aligning the relative Gibbs free energies per Na atom for NaBC and the three combinations of Na${_x}$BC with Na$_2^{\textrm{gas}}$. A similar procedure was done for Mg$_x$B$_2$C$_2$ using the oS48 and hP14 ($P6_3/mcm$) structures at $x=4/5$ and 2/3, respectively. The resulting phase diagrams show that deintercalation of the known MgB$_2$C$_2$ should well be achievable with similar experimental setups used for Li$_x$BC~\cite{Kharabadze2023, Zhao2003, Fogg2006}. NaBC may be desodiated under considerably lower temperatures. The large separation of stability regions suggests that attaining a certain $x$ composition in Na$_x$BC could be possible by choosing the targeted $(T,P_{\rm M})$ conditions and establishing the thermodynamic equilibrium, as opposed to relying upon the kinetics-determined non-equilibrium reaction~\cite{Kharabadze2023}.

\label{sec:Superconductivity}
\subsection{Superconductivity}

\begin{table*}[!htb]
\caption{Properties of select Na$_x$BC and Mg$_x$B$_2$C$_2$ phases: the thermodynamic stability ($\Delta E_{\rm hull}$) with respect to the local kinetically-restrained convex hull, average buckling of BC layers ($d_{\rm avg}$), density of states (DOS($E_{\rm F}$)), superconducting $\lambda$, and superconducting critical temperature calculated with the Allen-Dynes modified McMillian equation ($T_{\rm c}^{\rm AD}$) and with the aME formalism ($T_{\rm c}^{\rm aME}$).} \label{tab:T1}
\setlength\tabcolsep{0pt} 

\smallskip 
\begin{tabular*}{\textwidth}{@{\extracolsep{\fill}}l c c c c c  c c c }
\hline\hline \noalign{\vskip 1mm}
 Phase  & Space  & Pearson  & $\Delta E_{\rm hull}$ & $d_{\rm avg}$  & DOS($E_F$) &    $\lambda$   & $T_{\rm c}^{\rm AD}$   &  $T_{\rm c}^{\rm aME}$  \\
   composition    & group  & symbol   &  (meV/atom)             &  (\AA)            &   (states/(eV atom))          &                 &    (K)                 & (K) \\ \noalign{\vskip 1mm}
\midrule                      
Na$_{7/8}$BC           &   $P2/m$   & mP23  & 6.7  &  0.04 & 0.17         &  0.97       &   24.9                    &    88                      \\ 
Na$_{3/4}$BC           &   $Pbam$   & oP22  & 0.0  &  0.11 & 0.21         &  0.95       &    21.9                   &     84                     \\  
Na$_{2/3}$BC           &   $Cmc2_1$ & oS32  & 0.0  &  0.12 & 0.32          &  1.32       &    15.6               &     43                     \\  
Na$_{5/8}$BC           &   $P2/m$   & mP21  & 10.9 &  0.14 & 0.19          &  0.51       &     1.0                  &     -                     \smallskip \\   
Mg$_{4/5}$B$_2$C$_2$   &   $C222_1$          &  oS48 & 0.0 &  0.06 & 0.23     &  0.99       &   20.8                    &     57                     \\ 
Mg$_{3/4}$B$_2$C$_2$   &   $P2$              &  mP19 & 22.0 &  0.05 & 0.26    &  0.97       &   18.8                    &     59                     \\ 
Mg$_{2/3}$B$_2$C$_2$   &   $P\overline{6}2m$ &  hP7  & 0.0 &  0.00 & 0.27     &  1.12       &   28.7                    &     73                     \\ \noalign{\vskip 1mm} 
\hline\hline
\end{tabular*}
\end{table*}

The following analysis aims to establish whether the proposed layered Na$_x$BC and Mg$_x$B$_2$C$_2$ materials attainable at ambient pressure have the necessary electronic and vibrational properties to be high-$T_{\rm c}$ superconductors. Previous work~\cite{Rosner2002, Dewhurst2003, Tomassetti2024, Verma2003, Pham2023, Singh2002, Miao2016} has demonstrated similarities between hole-doped borocarbides and MgB$_2$, attributing the strong e-ph coupling to the pairing of hole-doped $\sigma$ states at the Fermi level with bond-stretching BC or B phonon modes. The superconducting properties of the relevant Mg$_x$B$_2$C$_2$, Na$_x$BC, and NaB$_{1+x}$C$_{1-x}$ compounds have been investigated in several studies~\cite{Verma2003, Pham2023, Singh2002, Miao2016} but the $T_{\rm c}$ estimates have not included the e-ph anisotropy important in MgB$_2$-type materials. The recent implementation of the anisotropic Migdal-Eliashberg formalism and the present identification of representative Na$_x$BC and Mg$_x$B$_2$C$_2$ materials allow us to make more accurate $T_{\rm c}$ predictions and examine the $T_{\rm c}$ sensitivity to different morphological traits.

\begin{figure*}[t!]
   \centering
\includegraphics[width=0.84\textwidth]{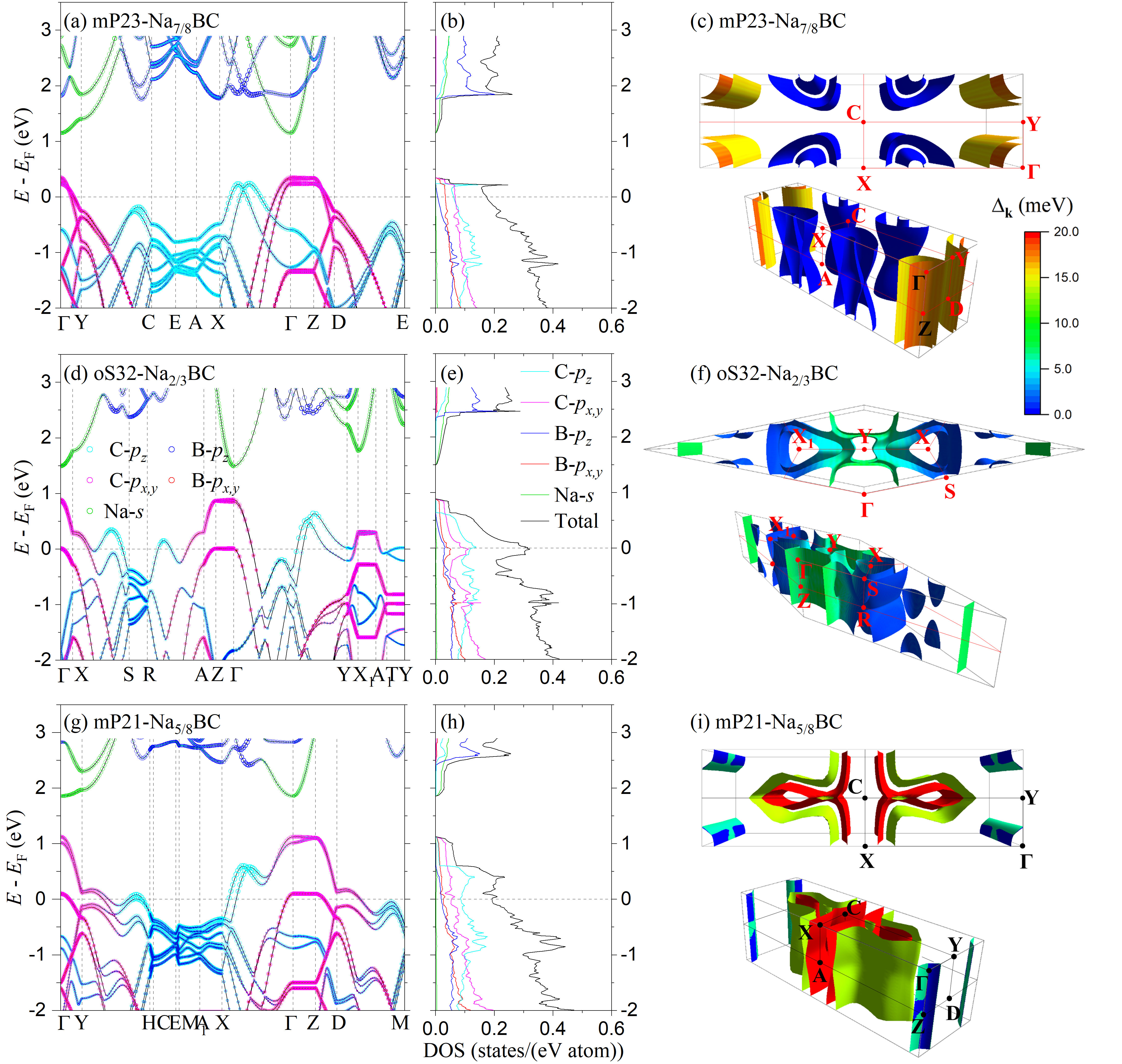}
    \caption{\label{fig-08} Electronic properties of (a),(b),(c) mP23-Na$_{7/8}$BC, (d),(e),(f) oS32-Na$_{2/3}$BC, and (g),(h),(i) mP21-Na$_{5/8}$BC. The left (middle) panels show band structures with orbital character (total and projected DOS). The right panels show Fermi surfaces where colors are used to show (c),(f) the value of the superconducting gap at 10 K or (i) the band index. mP23-Na$_{7/8}$BC possesses the signature MgB$_2$-like Fermi surface cylinders, while the other two phases with large BC layer buckling do not.}
\end{figure*}

First, we evaluated the potential for superconductivity in select phases depicted in Figs.~\ref{fig-01} and \ref{fig-04}, and listed in Table~\ref{tab:T1} to check for optimal doping levels in the Mg- and Na-based borocarbides. In addition to the honeycomb phases defining the local convex hull (oS48-Mg$_{4/5}$B$_2$C$_2$, hP14-Mg$_{2/3}$B$_2$C$_2$ ($P6_3/mcm$), oP22-Na$_{3/4}$BC and oS32-Na$_{2/3}$BC), the set included other low-energy metastable phases (mP19-Mg$_{3/4}$B$_2$C$_2$, hP7-Mg$_{2/3}$B$_2$C$_2$, mP23-Na$_{7/8}$BC, oP22-Na$_{3/4}$BC, and mP21-Na$_{5/8}$BC) to provide a better sampling of the composition range. All of the Mg borocarbides and the Na borocarbides with $x>2/3$ display the hallmark MgB$_2$ electronic properties, including the hole-doped B- and C-${\sigma}$ states lifted at $\Gamma$ by about 0.5 eV above the Fermi level (Figs.~\ref{fig-08}(a), S1 and S2~\cite{SM}), the corresponding sets of inner and outer Fermi surface cylinders (Figs.~S10 and S11~\cite{SM}), and a considerable total DOS of 0.23-0.27 (0.17-0.21) states/(eV atom) in the Mg (Na) phases (Figs.~\ref{fig-08}(b), S1 and S2~\cite{SM}).

To determine the importance of the stacking order of the BC sheets, we performed superconductivity calculations for the hP7 (AA) and hP14 (AA’, $P6_3/mcm$) Mg$_{2/3}$B$_2$C$_2$ variants, shown in Fig.~\ref{fig-04}(i)-(ii). The primary difference between them is a more pronounced dispersion of the C-$p_z$ states along the $\Gamma-A$ direction in the AA-stacked structure due to a larger interlayer overlap of the $p_z$ orbitals centered on the vertically aligned C atoms (Fig.~S3~\cite{SM}). While the stacking shift causes a significant rearrangement of the conduction states near 1.5~eV, it has a negligible effect on the total DOS($E_{\rm F}$), the isotropic e-ph coupling (Fig.~S6~\cite{SM}), or the anisotropic $T_{\rm c}$ (Fig.~S9~\cite{SM}). We checked the sensitivity of the results to the choice of the functional by examining hP7 in both optB86b and optB88 approximations. The latter produces a slightly larger 0.3\% interlayer spacing, a close in-plane lattice constant within 0.1\%, a 5 meV softening of the bond-stretching BC modes lying near 80 meV, an 8\% boost in the isotropic $\lambda$, and a 10~K increase in $T_{\rm c}$ (Figs.~S4, S6, S7 and S9~\cite{SM}).

The total e-ph coupling strengths in the majority of the examined layered borocarbides are near 1.0, but the Eliashberg spectral functions plotted in Figs.~S4 and S5~\cite{SM} help appreciate the different make-ups of the integral values. In the Mg-intercalated phases, only about half of the e-ph coupling comes from the BC bond-stretching phonon modes with frequencies above 75 meV, with the other half determined primarily by mixed soft modes below 50 meV. The soft modes were found to play a similarly important role in our study of the Li$_x$BC and Li$_x$M$_y$BC materials~\cite{Tomassetti2024}. In the Na borocarbides with $x = 7/8$ and $x = 3/4$, the bond-stretching modes soften significantly, down to 60 meV, and make a dominant contribution, between 70\% and 80\%, to the total coupling. In this respect, the two phases bear stronger resemblance to the MgB$_2$ superconductor~\cite{Kafle2022,Tomassetti2024}. Fittingly, calculations of the superconducting $T_{\rm c}$ with the anisotropic full-bandwidth Migdal-Eliashberg equations result in gap functions with a MgB$_2$-like two-gap structure, vanishing at 57~K, 59~K, 73~K, 88~K, and 84~K for the oS48-Mg$_{4/5}$B$_2$C$_2$, mP19-Mg$_{3/4}$B$_2$C$_2$, and hP7-Mg$_{2/3}$B$_2$C$_2$, mP23-Na$_{7/8}$BC, and oP22-Na$_{3/4}$BC phases, respectively (Figs. S7 and S8~\cite{SM}). 

The consistently high {\it ab initio} $T_{\rm c}$ values obtained for this set, as well as for a number of layered Li-Mg-B~\cite{Kafle2022} and Li$_x$M$_y$BC compounds~\cite{Tomassetti2024}, make it seem that any material with ordered hole-doped honeycomb frameworks should have robust phonon-mediated superconductivity, a notion not borne out by experiments on Li$_x$BC~\cite{Bharathi2002,  Nakamori2003, Zhao2003,Fogg2003a,Fogg2003b,Fogg2006, Kalkan2019} or Mg$_{0.5}$Li$_{0.8}$B$_2$C$_2$~\cite{Mori2004}. In addition to showing a $T_{\rm c}$ reduction caused by BC layer buckling and an effective superconductivity suppression caused by BC layer bridging or disorder, we use oS32-Na$_{2/3}$BC and mP21-Na$_{5/8}$BC as examples to further illustrate the detrimental effect relatively moderate deviations from the perfect planar morphology can have on the e-ph coupling.

At first glance, oS32-Na$_{2/3}$BC with a high 0.12-\AA\ average buckling appears to be an even better candidate for high-$T_{\rm c}$ superconductivity because of the highest DOS($E_{\rm F}$)=0.32 states/(eV atom) and $\lambda=1.32$ among the considered Na borocarbides (Figs.~\ref{fig-08}(e) and S5~\cite{SM}). However, the distribution of the electronic states and the breakdown of $\lambda$ contributions set it apart from the examined phases with higher Na content. Just as the curvature of C layers in nanotubes makes the valence states acquire a mixed $sp^2$/$sp^3$ hybrid character~\cite{dumitrica2002}, the distortion of the BC layers in this partially deintercalated metal borocarbide causes a consequential realignment of the C-$p$ states (we will refer to the in-plane and out-of-plane directions as $z$ and $(x,y)$ regardless of the orientation of the unit cell). Namely, a 0.5~eV gap opens up between the $p_z$ and $p_{x,y}$ states around $-0.5$~eV along the $R-A$ and $X-S$ directions, and the separation between the $p_{x,y}$ states along $\Gamma-Z$ increases to about 1~eV, which reshapes the outer Fermi surface cylinders from the upper $p_{x,y}$ states at 1 eV and collapses the inner ones from the lower sets that are now essentially at the Fermi level (Fig.~\ref{fig-08}(f)). As a result, the largest contribution to the e-ph coupling comes not from the bond-stretching BC phonon modes, but rather from low-frequency manifold with predominantly Na character (Fig.~S5~\cite{SM}). The superconducting gap function does not have the two typical distinct gaps (Fig.~S8~\cite{SM}) and the $T_{\rm c}$ can be expected to be around 43~K, but an unusually slow convergence in the iterative solution of the aME equations, presumably due to the presence of a sharp DOS peak at $E_{\rm F}$, introduces a higher degree of uncertainty in this estimate.

The mP21-Na$_{5/8}$BC phase with the largest $d_{\rm avg}=0.14$ \AA\ has a total DOS($E_{\rm F}$) value of 0.19 states/(eV atom) comparable to those in Na$_{x>2/3}$BC (Figs.~\ref{fig-08} and S2~\cite{SM}). The Eliashberg spectral function also has a well-defined maximum corresponding to the BC bond-stretching vibrations, but the peak is localized around 90~meV instead of spreading over a broad range and attaining its highest value at $\sim$65 meV. This indicates the lack of the characteristic phonon softening and leads to a net e-ph coupling value of only 0.51 (Fig.~S5~\cite{SM}). The band structure plot in Fig.~\ref{fig-08}(g) reveals that, just as in oS32-Na$_{2/3}$BC, the corrugation of the BC layer pushes the C-$p_{x,y}$ states further apart to 1 eV along $\Gamma-Z$. Despite the lower pair of the $p_{x,y}$ bands remaining hole-doped and generating the inner Fermi cylinders, the outer ones are no longer discernible (Fig.~\ref{fig-08}(i)). Given the low value of $\lambda$ and the 1~K estimate of the Allen-Dynes $T_{\rm c}$, we did not attempt aME calculations for this phase. 

Table~\ref{tab:T1} summarizes the aME results for Na and Mg borocarbides at all examined compositions. The $T_{\rm c}$ values in the Mg phases trend upward from 57~K to 73~K with decreasing $x$, closely following the aME estimates obtained for Li$_x$BC at the matching levels of the electron count~\cite{Tomassetti2024}. The highest $T_{\rm c}$ of 88~K for the Na phases, on the other hand, is found in the least hole-doped mP23-Na$_{7/8}$BC that lies above the local convex hull. The factor of two $T_{\rm c}$ reduction in oS32-Na$_{2/3}$BC with higher doped but significantly more puckered BC layers is in qualitative agreement with the trend found for Li$_{1/2}$BC~\cite{Tomassetti2024}. However, our findings for mP21-Na$_{5/8}$BC raise a question whether buckling alone could suppress superconductivity at other possible compositions or decorations not sampled in this study.

A systematic screening of superconducting properties at the highest level of theory would be difficult to carry out because of the prohibitively high computational cost of aME calculations. Several simple descriptors have been proposed in previous studies to detect specific signs of the MgB$_2$-type superconductivity. For example, the difference between $\Gamma$- and $M$-point bond-stretching phonon frequencies was found to be a suitable proxy for phonon softening in metal borides with hexagonal unit cells~\cite{ak14}, while a zone-center descriptor involving the comparison of screened phonon frequencies ($\omega$) obtained via standard calculations versus unscreened phonon frequencies ($\tilde{\omega}$) obtained by fixing the occupation number in the finite displacement method was shown to provide estimates of the e-ph coupling~\cite{descriptor-lambda}. Unfortunately, the application of the descriptors to the metal borocarbides proved to be not straightforward because of the large size and low symmetry of the considered structures and we opted for a manual inspection of the relevant electronic structure features. In order to distinguish between the electron count and BC layer corrugation factors, we focused on two representative compositions, Na$_{5/8}$BC and Mg$_{3/4}$B$_2$C$_2$, and considered an additional 8 Na and 7 Mg phases within 100 meV/atom from the respective most stable configuration. Upon examination of their band structures and Fermi surfaces, we confirm that all 3 Na and 4 Mg structures with the average buckling above 0.12 \AA\ have absent, distorted, or reduced Fermi surface cylinders (Figs.~S12 and S13~\cite{SM}). The results provide evidence that buckling is indeed a major factor capable of suppressing phonon-mediated superconductivity at various metal borocarbide stoichiometries.

\section{Conclusions}
\label{sec:conclusions}

Our {\it ab initio} investigation into the stability of Na and Mg borocarbides uncovers layered oP22-Na$_{3/4}$BC, oS32-Na$_{2/3}$BC, oS80-Na$_{1/2}$BC, oS48-Mg$_{4/5}$B$_2$C$_2$, and hP14-Mg$_{2/3}$B$_2$C$_2$ ($P6_3/mcm$) phases as the most likely products of high-temperature deintercalation of the respective fully-loaded NaBC and MgB$_2$C$_2$ materials. The proposed ($T,P_{\rm M}$) diagrams establish temperature and Na$_2$ (Mg) gas partial pressure conditions defining phase boundaries between the metal borocarbides as long as they retain the layered BC morphology. The deintercalation of NaBC requires much lower temperatures and may produce Na$_x$BC derivatives with fewer defects in the BC honeycomb layers. Examination of layered Na and Mg borocarbides at various compositions with aME reveals that buckling of the covalent honeycomb BC networks, that tends to occur in heavily deintercalated Na$_{2/3>x}$BC, can effectively suppress MgB$_2$-type superconductivity. Fortunately, our aME calculations indicate that all {\it accessible} hole-doped phases identified in this study should have high $T_{\rm c}$ between 43~K and 84~K. We hope that the combination of the extensive experimental knowledge of the related Li and Mg borocarbides and the present {\it ab initio} findings can guide the synthesis of new ambient-pressure high-$T_{\rm c}$ superconductors.

\begin{acknowledgments} 
The authors acknowledge support from the National Science Foundation (NSF) (Award No. DMR-2320073). This work used the Frontera supercomputer at the Texas Advanced Computing Center via the Leadership Resource Allocation (LRAC) award DMR22004 and the Expanse system at the San Diego Supercomputer Center through the NSF-supported ACCESS program (allocation TG-DMR180071). 


\end{acknowledgments}

\bibliography{paper} 

\begin{thebibliography}{74}%
\makeatletter
\providecommand \@ifxundefined [1]{%
 \@ifx{#1\undefined}
}%
\providecommand \@ifnum [1]{%
 \ifnum #1\expandafter \@firstoftwo
 \else \expandafter \@secondoftwo
 \fi
}%
\providecommand \@ifx [1]{%
 \ifx #1\expandafter \@firstoftwo
 \else \expandafter \@secondoftwo
 \fi
}%
\providecommand \natexlab [1]{#1}%
\providecommand \enquote  [1]{``#1''}%
\providecommand \bibnamefont  [1]{#1}%
\providecommand \bibfnamefont [1]{#1}%
\providecommand \citenamefont [1]{#1}%
\providecommand \href@noop [0]{\@secondoftwo}%
\providecommand \href [0]{\begingroup \@sanitize@url \@href}%
\providecommand \@href[1]{\@@startlink{#1}\@@href}%
\providecommand \@@href[1]{\endgroup#1\@@endlink}%
\providecommand \@sanitize@url [0]{\catcode `\\12\catcode `\$12\catcode
  `\&12\catcode `\#12\catcode `\^12\catcode `\_12\catcode `\%12\relax}%
\providecommand \@@startlink[1]{}%
\providecommand \@@endlink[0]{}%
\providecommand \url  [0]{\begingroup\@sanitize@url \@url }%
\providecommand \@url [1]{\endgroup\@href {#1}{\urlprefix }}%
\providecommand \urlprefix  [0]{URL }%
\providecommand \Eprint [0]{\href }%
\providecommand \doibase [0]{https://doi.org/}%
\providecommand \selectlanguage [0]{\@gobble}%
\providecommand \bibinfo  [0]{\@secondoftwo}%
\providecommand \bibfield  [0]{\@secondoftwo}%
\providecommand \translation [1]{[#1]}%
\providecommand \BibitemOpen [0]{}%
\providecommand \bibitemStop [0]{}%
\providecommand \bibitemNoStop [0]{.\EOS\space}%
\providecommand \EOS [0]{\spacefactor3000\relax}%
\providecommand \BibitemShut  [1]{\csname bibitem#1\endcsname}%
\let\auto@bib@innerbib\@empty
\bibitem [{\citenamefont {Nagamatsu}\ \emph {et~al.}(2001)\citenamefont
  {Nagamatsu}, \citenamefont {Nakagawa}, \citenamefont {Muranaka},
  \citenamefont {Zenitani},\ and\ \citenamefont {Akimitsu}}]{Nagamatsu2001}%
  \BibitemOpen
  \bibfield  {author} {\bibinfo {author} {\bibfnamefont {J.}~\bibnamefont
  {Nagamatsu}}, \bibinfo {author} {\bibfnamefont {N.}~\bibnamefont {Nakagawa}},
  \bibinfo {author} {\bibfnamefont {T.}~\bibnamefont {Muranaka}}, \bibinfo
  {author} {\bibfnamefont {Y.}~\bibnamefont {Zenitani}},\ and\ \bibinfo
  {author} {\bibfnamefont {J.}~\bibnamefont {Akimitsu}},\ }\bibfield  {title}
  {\bibinfo {title} {Superconductivity at 39 {K} in magnesium diboride},\
  }\href {http://www.nature.com/articles/35065039} {\bibfield  {journal}
  {\bibinfo  {journal} {Nature}\ }\textbf {\bibinfo {volume} {410}},\ \bibinfo
  {pages} {63} (\bibinfo {year} {2001})}\BibitemShut {NoStop}%
\bibitem [{\citenamefont {Pelleg}\ \emph {et~al.}(2007)\citenamefont {Pelleg},
  \citenamefont {Rotman},\ and\ \citenamefont {Sinder}}]{Pelleg2007}%
  \BibitemOpen
  \bibfield  {author} {\bibinfo {author} {\bibfnamefont {J.}~\bibnamefont
  {Pelleg}}, \bibinfo {author} {\bibfnamefont {M.}~\bibnamefont {Rotman}},\
  and\ \bibinfo {author} {\bibfnamefont {M.}~\bibnamefont {Sinder}},\
  }\bibfield  {title} {\bibinfo {title} {{Borides of Ag and Au prepared by
  magnetron sputtering}},\ }\href {https://doi.org/10.1016/j.physc.2007.06.009}
  {\bibfield  {journal} {\bibinfo  {journal} {Physica C: Superconductivity}\
  }\textbf {\bibinfo {volume} {466}},\ \bibinfo {pages} {61} (\bibinfo {year}
  {2007})}\BibitemShut {NoStop}%
\bibitem [{\citenamefont {Kolmogorov}\ and\ \citenamefont
  {Curtarolo}(2006{\natexlab{a}})}]{ak09}%
  \BibitemOpen
  \bibfield  {author} {\bibinfo {author} {\bibfnamefont {A.~N.}\ \bibnamefont
  {Kolmogorov}}\ and\ \bibinfo {author} {\bibfnamefont {S.}~\bibnamefont
  {Curtarolo}},\ }\bibfield  {title} {\bibinfo {title} {Theoretical study of
  metal borides stability},\ }\href
  {https://doi.org/10.1103/PhysRevB.74.224507} {\bibfield  {journal} {\bibinfo
  {journal} {Physical Review B}\ }\textbf {\bibinfo {volume} {74}},\ \bibinfo
  {pages} {224507} (\bibinfo {year} {2006}{\natexlab{a}})}\BibitemShut
  {NoStop}%
\bibitem [{\citenamefont {Kolmogorov}\ and\ \citenamefont
  {Curtarolo}(2006{\natexlab{b}})}]{ak08}%
  \BibitemOpen
  \bibfield  {author} {\bibinfo {author} {\bibfnamefont {A.~N.}\ \bibnamefont
  {Kolmogorov}}\ and\ \bibinfo {author} {\bibfnamefont {S.}~\bibnamefont
  {Curtarolo}},\ }\bibfield  {title} {\bibinfo {title} {{Prediction of
  different crystal structure phases in metal borides: A lithium monoboride
  analog to MgB$_2$}},\ }\href {https://doi.org/10.1103/PhysRevB.73.180501}
  {\bibfield  {journal} {\bibinfo  {journal} {Physical Review B}\ }\textbf
  {\bibinfo {volume} {73}},\ \bibinfo {pages} {180501} (\bibinfo {year}
  {2006}{\natexlab{b}})}\BibitemShut {NoStop}%
\bibitem [{\citenamefont {Calandra}\ \emph {et~al.}(2007)\citenamefont
  {Calandra}, \citenamefont {Kolmogorov},\ and\ \citenamefont
  {Curtarolo}}]{ak10}%
  \BibitemOpen
  \bibfield  {author} {\bibinfo {author} {\bibfnamefont {M.}~\bibnamefont
  {Calandra}}, \bibinfo {author} {\bibfnamefont {A.~N.}\ \bibnamefont
  {Kolmogorov}},\ and\ \bibinfo {author} {\bibfnamefont {S.}~\bibnamefont
  {Curtarolo}},\ }\bibfield  {title} {\bibinfo {title} {{Search for high $T_c$
  in layered structures: The case of LiB}},\ }\href
  {https://doi.org/10.1103/PhysRevB.75.144506} {\bibfield  {journal} {\bibinfo
  {journal} {Physical Review B}\ }\textbf {\bibinfo {volume} {75}},\ \bibinfo
  {pages} {144506} (\bibinfo {year} {2007})}\BibitemShut {NoStop}%
\bibitem [{\citenamefont {Kolmogorov}\ \emph {et~al.}(2008)\citenamefont
  {Kolmogorov}, \citenamefont {Calandra},\ and\ \citenamefont
  {Curtarolo}}]{ak14}%
  \BibitemOpen
  \bibfield  {author} {\bibinfo {author} {\bibfnamefont {A.~N.}\ \bibnamefont
  {Kolmogorov}}, \bibinfo {author} {\bibfnamefont {M.}~\bibnamefont
  {Calandra}},\ and\ \bibinfo {author} {\bibfnamefont {S.}~\bibnamefont
  {Curtarolo}},\ }\bibfield  {title} {\bibinfo {title} {{Thermodynamic
  stabilities of ternary metal borides: An \it{ab} \it{initio} \rm guide for
  synthesizing layered superconductors}},\ }\href
  {https://doi.org/10.1103/PhysRevB.78.094520} {\bibfield  {journal} {\bibinfo
  {journal} {Physical Review B}\ }\textbf {\bibinfo {volume} {78}},\ \bibinfo
  {pages} {094520} (\bibinfo {year} {2008})}\BibitemShut {NoStop}%
\bibitem [{\citenamefont {Kolmogorov}\ \emph {et~al.}(2015)\citenamefont
  {Kolmogorov}, \citenamefont {Hajinazar}, \citenamefont {Angyal},
  \citenamefont {Kuznetsov},\ and\ \citenamefont {Jephcoat}}]{ak30}%
  \BibitemOpen
  \bibfield  {author} {\bibinfo {author} {\bibfnamefont {A.~N.}\ \bibnamefont
  {Kolmogorov}}, \bibinfo {author} {\bibfnamefont {S.}~\bibnamefont
  {Hajinazar}}, \bibinfo {author} {\bibfnamefont {C.}~\bibnamefont {Angyal}},
  \bibinfo {author} {\bibfnamefont {V.~L.}\ \bibnamefont {Kuznetsov}},\ and\
  \bibinfo {author} {\bibfnamefont {A.~P.}\ \bibnamefont {Jephcoat}},\
  }\bibfield  {title} {\bibinfo {title} {Synthesis of a predicted layered {LiB}
  via cold compression},\ }\href
  {https://link.aps.org/doi/10.1103/PhysRevB.92.144110} {\bibfield  {journal}
  {\bibinfo  {journal} {Phys. Rev. B}\ }\textbf {\bibinfo {volume} {92}},\
  \bibinfo {pages} {144110} (\bibinfo {year} {2015})}\BibitemShut {NoStop}%
\bibitem [{\citenamefont {Cava}\ \emph {et~al.}(2003)\citenamefont {Cava},
  \citenamefont {Zandbergen},\ and\ \citenamefont {Inumaru}}]{Cava2003}%
  \BibitemOpen
  \bibfield  {author} {\bibinfo {author} {\bibfnamefont {R.~J.}\ \bibnamefont
  {Cava}}, \bibinfo {author} {\bibfnamefont {H.~W.}\ \bibnamefont
  {Zandbergen}},\ and\ \bibinfo {author} {\bibfnamefont {K.}~\bibnamefont
  {Inumaru}},\ }\bibfield  {title} {\bibinfo {title} {{The substitutional
  chemistry of MgB$_2$}},\ }\href
  {https://doi.org/10.1016/S0921-4534(02)02327-4} {\bibfield  {journal}
  {\bibinfo  {journal} {Physica C: Supercond.}\ }\textbf {\bibinfo {volume}
  {385}},\ \bibinfo {pages} {8} (\bibinfo {year} {2003})}\BibitemShut {NoStop}%
\bibitem [{\citenamefont {Bianconi}\ \emph {et~al.}(2007)\citenamefont
  {Bianconi}, \citenamefont {Busby}, \citenamefont {Fratini}, \citenamefont
  {Palmisano}, \citenamefont {Simonelli}, \citenamefont {Filippi},
  \citenamefont {Sanna}, \citenamefont {Congiu}, \citenamefont {Saccone},
  \citenamefont {Giovannini},\ and\ \citenamefont {{De Negri}}}]{Bianconi2007}%
  \BibitemOpen
  \bibfield  {author} {\bibinfo {author} {\bibfnamefont {A.}~\bibnamefont
  {Bianconi}}, \bibinfo {author} {\bibfnamefont {Y.}~\bibnamefont {Busby}},
  \bibinfo {author} {\bibfnamefont {M.}~\bibnamefont {Fratini}}, \bibinfo
  {author} {\bibfnamefont {V.}~\bibnamefont {Palmisano}}, \bibinfo {author}
  {\bibfnamefont {L.}~\bibnamefont {Simonelli}}, \bibinfo {author}
  {\bibfnamefont {M.}~\bibnamefont {Filippi}}, \bibinfo {author} {\bibfnamefont
  {S.}~\bibnamefont {Sanna}}, \bibinfo {author} {\bibfnamefont
  {F.}~\bibnamefont {Congiu}}, \bibinfo {author} {\bibfnamefont
  {A.}~\bibnamefont {Saccone}}, \bibinfo {author} {\bibfnamefont
  {M.}~\bibnamefont {Giovannini}},\ and\ \bibinfo {author} {\bibfnamefont
  {S.}~\bibnamefont {{De Negri}}},\ }\bibfield  {title} {\bibinfo {title}
  {{Controlling the Critical Temperature in Mg$_{1-x}$Al$_x$B$_2$}},\ }\href
  {http://link.springer.com/10.1007/s10948-007-0279-7} {\bibfield  {journal}
  {\bibinfo  {journal} {J. Supercond. Nov. Magn.}\ }\textbf {\bibinfo {volume}
  {20}},\ \bibinfo {pages} {495} (\bibinfo {year} {2007})}\BibitemShut
  {NoStop}%
\bibitem [{\citenamefont {Karpinski}\ \emph {et~al.}(2008)\citenamefont
  {Karpinski}, \citenamefont {Zhigadlo}, \citenamefont {Katrych}, \citenamefont
  {Rogacki}, \citenamefont {Batlogg}, \citenamefont {Tortello},\ and\
  \citenamefont {Puzniak}}]{Karpinski2008}%
  \BibitemOpen
  \bibfield  {author} {\bibinfo {author} {\bibfnamefont {J.}~\bibnamefont
  {Karpinski}}, \bibinfo {author} {\bibfnamefont {N.~D.}\ \bibnamefont
  {Zhigadlo}}, \bibinfo {author} {\bibfnamefont {S.}~\bibnamefont {Katrych}},
  \bibinfo {author} {\bibfnamefont {K.}~\bibnamefont {Rogacki}}, \bibinfo
  {author} {\bibfnamefont {B.}~\bibnamefont {Batlogg}}, \bibinfo {author}
  {\bibfnamefont {M.}~\bibnamefont {Tortello}},\ and\ \bibinfo {author}
  {\bibfnamefont {R.}~\bibnamefont {Puzniak}},\ }\bibfield  {title} {\bibinfo
  {title} {{MgB$_2$ single crystals substituted with Li and with Li-C:
  Structural and superconducting properties}},\ }\href
  {https://doi.org/10.1103/PhysRevB.77.214507} {\bibfield  {journal} {\bibinfo
  {journal} {Phys. Rev. B}\ }\textbf {\bibinfo {volume} {77}},\ \bibinfo
  {pages} {214507} (\bibinfo {year} {2008})}\BibitemShut {NoStop}%
\bibitem [{\citenamefont {Parisiades}\ \emph {et~al.}(2009)\citenamefont
  {Parisiades}, \citenamefont {Liarokapis}, \citenamefont {Zhigadlo},
  \citenamefont {Katrych},\ and\ \citenamefont {Karpinski}}]{Parisiades2009}%
  \BibitemOpen
  \bibfield  {author} {\bibinfo {author} {\bibfnamefont {P.}~\bibnamefont
  {Parisiades}}, \bibinfo {author} {\bibfnamefont {E.}~\bibnamefont
  {Liarokapis}}, \bibinfo {author} {\bibfnamefont {N.~D.}\ \bibnamefont
  {Zhigadlo}}, \bibinfo {author} {\bibfnamefont {S.}~\bibnamefont {Katrych}},\
  and\ \bibinfo {author} {\bibfnamefont {J.}~\bibnamefont {Karpinski}},\
  }\bibfield  {title} {\bibinfo {title} {{Raman Investigations of C-, Li- and
  Mn-Doped MgB$_2$}},\ }\href {https://doi.org/10.1007/s10948-008-0402-4}
  {\bibfield  {journal} {\bibinfo  {journal} {J. Supercond. Nov. Magn.}\
  }\textbf {\bibinfo {volume} {22}},\ \bibinfo {pages} {169} (\bibinfo {year}
  {2009})}\BibitemShut {NoStop}%
\bibitem [{\citenamefont {Calandra}\ and\ \citenamefont
  {Mauri}(2005)}]{Calandra2005}%
  \BibitemOpen
  \bibfield  {author} {\bibinfo {author} {\bibfnamefont {M.}~\bibnamefont
  {Calandra}}\ and\ \bibinfo {author} {\bibfnamefont {F.}~\bibnamefont
  {Mauri}},\ }\bibfield  {title} {\bibinfo {title} {{Theoretical Explanation of
  Superconductivity in C$_6$Ca}},\ }\href
  {https://doi.org/10.1103/PhysRevLett.95.237002} {\bibfield  {journal}
  {\bibinfo  {journal} {Physical Review Letters}\ }\textbf {\bibinfo {volume}
  {95}},\ \bibinfo {pages} {237002} (\bibinfo {year} {2005})}\BibitemShut
  {NoStop}%
\bibitem [{\citenamefont {Mazin}(2005)}]{Mazin2005}%
  \BibitemOpen
  \bibfield  {author} {\bibinfo {author} {\bibfnamefont {I.}~\bibnamefont
  {Mazin}},\ }\bibfield  {title} {\bibinfo {title} {{Intercalant-Driven
  Superconductivity in YbC$_6$ and CaC$_6$}},\ }\bibfield  {journal} {\bibinfo
  {journal} {Physical Review Letters}\ }\textbf {\bibinfo {volume} {20375}},\
  \href {https://doi.org/10.1103/PhysRevLett.95.227001}
  {10.1103/PhysRevLett.95.227001} (\bibinfo {year} {2005})\BibitemShut
  {NoStop}%
\bibitem [{\citenamefont {Margine}\ \emph {et~al.}(2016)\citenamefont
  {Margine}, \citenamefont {Lambert},\ and\ \citenamefont
  {Giustino}}]{Margine2016}%
  \BibitemOpen
  \bibfield  {author} {\bibinfo {author} {\bibfnamefont {E.~R.}\ \bibnamefont
  {Margine}}, \bibinfo {author} {\bibfnamefont {H.}~\bibnamefont {Lambert}},\
  and\ \bibinfo {author} {\bibfnamefont {F.}~\bibnamefont {Giustino}},\
  }\bibfield  {title} {\bibinfo {title} {{Electron-phonon interaction and
  pairing mechanism in superconducting Ca-intercalated bilayer graphene}},\
  }\bibfield  {journal} {\bibinfo  {journal} {Scientific Reports}\ }\textbf
  {\bibinfo {volume} {6}},\ \href {https://doi.org/10.1038/srep21414}
  {10.1038/srep21414} (\bibinfo {year} {2016})\BibitemShut {NoStop}%
\bibitem [{\citenamefont {W{\"o}rle}\ \emph {et~al.}(1995)\citenamefont
  {W{\"o}rle}, \citenamefont {Nesper}, \citenamefont {Mair}, \citenamefont
  {Schwarz},\ and\ \citenamefont {Von~Schnering}}]{worle1995libc}%
  \BibitemOpen
  \bibfield  {author} {\bibinfo {author} {\bibfnamefont {M.}~\bibnamefont
  {W{\"o}rle}}, \bibinfo {author} {\bibfnamefont {R.}~\bibnamefont {Nesper}},
  \bibinfo {author} {\bibfnamefont {G.}~\bibnamefont {Mair}}, \bibinfo {author}
  {\bibfnamefont {M.}~\bibnamefont {Schwarz}},\ and\ \bibinfo {author}
  {\bibfnamefont {H.~G.}\ \bibnamefont {Von~Schnering}},\ }\bibfield  {title}
  {\bibinfo {title} {{LiBC}—ein vollst{\"a}ndig interkalierter
  heterographit},\ }\href {https://doi.org/10.1002/zaac.19956210707} {\bibfield
   {journal} {\bibinfo  {journal} {Z. fur Anorg. Allg. Chem.}\ }\textbf
  {\bibinfo {volume} {621}},\ \bibinfo {pages} {1153} (\bibinfo {year}
  {1995})}\BibitemShut {NoStop}%
\bibitem [{\citenamefont {W{\"o}rle}\ and\ \citenamefont
  {Nesper}(1994)}]{worle1994}%
  \BibitemOpen
  \bibfield  {author} {\bibinfo {author} {\bibfnamefont {M.}~\bibnamefont
  {W{\"o}rle}}\ and\ \bibinfo {author} {\bibfnamefont {R.}~\bibnamefont
  {Nesper}},\ }\bibfield  {title} {\bibinfo {title} {{MgB$_2$C$_2$}, a new
  graphite-related refractory compound},\ }\href
  {https://doi.org/https://doi.org/10.1016/0925-8388(94)91045-6} {\bibfield
  {journal} {\bibinfo  {journal} {J. Alloys Compd.}\ }\textbf {\bibinfo
  {volume} {216}},\ \bibinfo {pages} {75} (\bibinfo {year} {1994})}\BibitemShut
  {NoStop}%
\bibitem [{\citenamefont {Ravindran}\ \emph {et~al.}(2001)\citenamefont
  {Ravindran}, \citenamefont {Vajeeston}, \citenamefont {Vidya}, \citenamefont
  {Kjekshus},\ and\ \citenamefont {Fjellvåg}}]{Ravindran2001}%
  \BibitemOpen
  \bibfield  {author} {\bibinfo {author} {\bibfnamefont {P.}~\bibnamefont
  {Ravindran}}, \bibinfo {author} {\bibfnamefont {P.}~\bibnamefont
  {Vajeeston}}, \bibinfo {author} {\bibfnamefont {R.}~\bibnamefont {Vidya}},
  \bibinfo {author} {\bibfnamefont {A.}~\bibnamefont {Kjekshus}},\ and\
  \bibinfo {author} {\bibfnamefont {H.}~\bibnamefont {Fjellvåg}},\ }\bibfield
  {title} {\bibinfo {title} {{Detailed electronic structure studies on
  superconducting MgB$_2$ and related compounds}},\ }\href
  {https://doi.org/10.1103/PhysRevB.64.224509} {\bibfield  {journal} {\bibinfo
  {journal} {Phys. Rev. B}\ }\textbf {\bibinfo {volume} {64}},\ \bibinfo
  {pages} {224509} (\bibinfo {year} {2001})}\BibitemShut {NoStop}%
\bibitem [{\citenamefont {Rosner}\ \emph {et~al.}(2002)\citenamefont {Rosner},
  \citenamefont {Kitaigorodsky},\ and\ \citenamefont {Pickett}}]{Rosner2002}%
  \BibitemOpen
  \bibfield  {author} {\bibinfo {author} {\bibfnamefont {H.}~\bibnamefont
  {Rosner}}, \bibinfo {author} {\bibfnamefont {A.}~\bibnamefont
  {Kitaigorodsky}},\ and\ \bibinfo {author} {\bibfnamefont {W.~E.}\
  \bibnamefont {Pickett}},\ }\bibfield  {title} {\bibinfo {title} {Prediction
  of high {$T_{\rm c}$} superconductivity in hole-doped {LiBC}},\ }\href
  {https://doi.org/10.1103/PhysRevLett.88.127001} {\bibfield  {journal}
  {\bibinfo  {journal} {Phys. Rev. Lett.}\ }\textbf {\bibinfo {volume} {88}},\
  \bibinfo {pages} {127001} (\bibinfo {year} {2002})}\BibitemShut {NoStop}%
\bibitem [{\citenamefont {Dewhurst}\ \emph {et~al.}(2003)\citenamefont
  {Dewhurst}, \citenamefont {Sharma}, \citenamefont {Ambrosch-Draxl},\ and\
  \citenamefont {Johansson}}]{Dewhurst2003}%
  \BibitemOpen
  \bibfield  {author} {\bibinfo {author} {\bibfnamefont {J.~K.}\ \bibnamefont
  {Dewhurst}}, \bibinfo {author} {\bibfnamefont {S.}~\bibnamefont {Sharma}},
  \bibinfo {author} {\bibfnamefont {C.}~\bibnamefont {Ambrosch-Draxl}},\ and\
  \bibinfo {author} {\bibfnamefont {B.}~\bibnamefont {Johansson}},\ }\bibfield
  {title} {\bibinfo {title} {First-principles calculation of superconductivity
  in hole-doped {LiBC}: {$T_{\rm c}$ 65~K}},\ }\href
  {https://doi.org/10.1103/PhysRevB.68.020504} {\bibfield  {journal} {\bibinfo
  {journal} {Phys. Rev. B}\ }\textbf {\bibinfo {volume} {68}},\ \bibinfo
  {pages} {020504} (\bibinfo {year} {2003})}\BibitemShut {NoStop}%
\bibitem [{\citenamefont {Bharathi}\ \emph {et~al.}(2002)\citenamefont
  {Bharathi}, \citenamefont {Balaselvi}, \citenamefont {Premila}, \citenamefont
  {Sairam}, \citenamefont {Reddy}, \citenamefont {Sundar},\ and\ \citenamefont
  {Hariharan}}]{Bharathi2002}%
  \BibitemOpen
  \bibfield  {author} {\bibinfo {author} {\bibfnamefont {A.}~\bibnamefont
  {Bharathi}}, \bibinfo {author} {\bibfnamefont {S.~J.}\ \bibnamefont
  {Balaselvi}}, \bibinfo {author} {\bibfnamefont {M.}~\bibnamefont {Premila}},
  \bibinfo {author} {\bibfnamefont {T.}~\bibnamefont {Sairam}}, \bibinfo
  {author} {\bibfnamefont {G.}~\bibnamefont {Reddy}}, \bibinfo {author}
  {\bibfnamefont {C.}~\bibnamefont {Sundar}},\ and\ \bibinfo {author}
  {\bibfnamefont {Y.}~\bibnamefont {Hariharan}},\ }\bibfield  {title} {\bibinfo
  {title} {{Synthesis and search for superconductivity in LiBC}},\ }\href
  {https://doi.org/10.1016/S0038-1098(02)00474-X} {\bibfield  {journal}
  {\bibinfo  {journal} {Solid State Commun.}\ }\textbf {\bibinfo {volume}
  {124}},\ \bibinfo {pages} {423} (\bibinfo {year} {2002})}\BibitemShut
  {NoStop}%
\bibitem [{\citenamefont {Nakamori}\ and\ \citenamefont {ichi
  Orimo}(2003)}]{Nakamori2003}%
  \BibitemOpen
  \bibfield  {author} {\bibinfo {author} {\bibfnamefont {Y.}~\bibnamefont
  {Nakamori}}\ and\ \bibinfo {author} {\bibfnamefont {S.}~\bibnamefont {ichi
  Orimo}},\ }\bibfield  {title} {\bibinfo {title} {{Synthesis and
  characterization of single phase Li$_x$BC ($x$=0.5 and 1.0) using Li hydride
  as a starting material}},\ }\href
  {https://doi.org/10.1016/j.jallcom.2003.09.034} {\bibfield  {journal}
  {\bibinfo  {journal} {J. Alloys Compd.}\ }\textbf {\bibinfo {volume} {370}},\
  \bibinfo {pages} {L7} (\bibinfo {year} {2003})}\BibitemShut {NoStop}%
\bibitem [{\citenamefont {Zhao}\ \emph {et~al.}(2003)\citenamefont {Zhao},
  \citenamefont {Klavins},\ and\ \citenamefont {Liu}}]{Zhao2003}%
  \BibitemOpen
  \bibfield  {author} {\bibinfo {author} {\bibfnamefont {L.}~\bibnamefont
  {Zhao}}, \bibinfo {author} {\bibfnamefont {P.}~\bibnamefont {Klavins}},\ and\
  \bibinfo {author} {\bibfnamefont {K.}~\bibnamefont {Liu}},\ }\bibfield
  {title} {\bibinfo {title} {Synthesis and properties of hole-doped
  {Li$_{1-x}$BC}},\ }\href {https://doi.org/10.1063/1.1556285} {\bibfield
  {journal} {\bibinfo  {journal} {J. Appl. Phys.}\ }\textbf {\bibinfo {volume}
  {93}},\ \bibinfo {pages} {8653} (\bibinfo {year} {2003})}\BibitemShut
  {NoStop}%
\bibitem [{\citenamefont {Fogg}\ \emph
  {et~al.}(2003{\natexlab{a}})\citenamefont {Fogg}, \citenamefont {Claridge},
  \citenamefont {Darling},\ and\ \citenamefont {Rosseinsky}}]{Fogg2003a}%
  \BibitemOpen
  \bibfield  {author} {\bibinfo {author} {\bibfnamefont {A.~M.}\ \bibnamefont
  {Fogg}}, \bibinfo {author} {\bibfnamefont {J.~B.}\ \bibnamefont {Claridge}},
  \bibinfo {author} {\bibfnamefont {G.~R.}\ \bibnamefont {Darling}},\ and\
  \bibinfo {author} {\bibfnamefont {M.~J.}\ \bibnamefont {Rosseinsky}},\
  }\bibfield  {title} {\bibinfo {title} {Synthesis and characteristion of
  {Li$_x$BC}—hole doping does not induce superconductivity},\ }\href
  {https://doi.org/10.1039/B302058D} {\bibfield  {journal} {\bibinfo  {journal}
  {Chem. Commun.}\ }\textbf {\bibinfo {volume} {3}},\ \bibinfo {pages} {1348}
  (\bibinfo {year} {2003}{\natexlab{a}})}\BibitemShut {NoStop}%
\bibitem [{\citenamefont {Fogg}\ \emph
  {et~al.}(2003{\natexlab{b}})\citenamefont {Fogg}, \citenamefont {Chalker},
  \citenamefont {Claridge}, \citenamefont {Darling},\ and\ \citenamefont
  {Rosseinsky}}]{Fogg2003b}%
  \BibitemOpen
  \bibfield  {author} {\bibinfo {author} {\bibfnamefont {A.~M.}\ \bibnamefont
  {Fogg}}, \bibinfo {author} {\bibfnamefont {P.~R.}\ \bibnamefont {Chalker}},
  \bibinfo {author} {\bibfnamefont {J.~B.}\ \bibnamefont {Claridge}}, \bibinfo
  {author} {\bibfnamefont {G.~R.}\ \bibnamefont {Darling}},\ and\ \bibinfo
  {author} {\bibfnamefont {M.~J.}\ \bibnamefont {Rosseinsky}},\ }\bibfield
  {title} {\bibinfo {title} {{LiBC} electronic, vibrational, structural, and
  low-temperature chemical behavior of a layered material isoelectronic with
  {MgB$_2$}},\ }\href {https://doi.org/10.1103/PhysRevB.67.245106} {\bibfield
  {journal} {\bibinfo  {journal} {Phys. Rev. B}\ }\textbf {\bibinfo {volume}
  {67}},\ \bibinfo {pages} {245106} (\bibinfo {year}
  {2003}{\natexlab{b}})}\BibitemShut {NoStop}%
\bibitem [{\citenamefont {Fogg}\ \emph {et~al.}(2006)\citenamefont {Fogg},
  \citenamefont {Meldrum}, \citenamefont {Darling}, \citenamefont {Claridge},\
  and\ \citenamefont {Rosseinsky}}]{Fogg2006}%
  \BibitemOpen
  \bibfield  {author} {\bibinfo {author} {\bibfnamefont {A.~M.}\ \bibnamefont
  {Fogg}}, \bibinfo {author} {\bibfnamefont {J.}~\bibnamefont {Meldrum}},
  \bibinfo {author} {\bibfnamefont {G.~R.}\ \bibnamefont {Darling}}, \bibinfo
  {author} {\bibfnamefont {J.~B.}\ \bibnamefont {Claridge}},\ and\ \bibinfo
  {author} {\bibfnamefont {M.~J.}\ \bibnamefont {Rosseinsky}},\ }\bibfield
  {title} {\bibinfo {title} {{Chemical control of electronic structure and
  superconductivity in layered borides and borocarbides: understanding the
  absence of superconductivity in {Li$_x$BC}}},\ }\href
  {https://pubs.acs.org/doi/10.1021/ja0578449} {\bibfield  {journal} {\bibinfo
  {journal} {J. Am. Chem. Soc.}\ }\textbf {\bibinfo {volume} {128}},\ \bibinfo
  {pages} {10043} (\bibinfo {year} {2006})}\BibitemShut {NoStop}%
\bibitem [{\citenamefont {Kalkan}\ and\ \citenamefont
  {Ozdas}(2019)}]{Kalkan2019}%
  \BibitemOpen
  \bibfield  {author} {\bibinfo {author} {\bibfnamefont {B.}~\bibnamefont
  {Kalkan}}\ and\ \bibinfo {author} {\bibfnamefont {E.}~\bibnamefont {Ozdas}},\
  }\bibfield  {title} {\bibinfo {title} {Staging phenomena in
  lithium-intercalated boron–carbon},\ }\href
  {https://doi.org/10.1021/acsami.8b19142} {\bibfield  {journal} {\bibinfo
  {journal} {ACS Appl. Mater. Interfaces}\ }\textbf {\bibinfo {volume} {11}},\
  \bibinfo {pages} {4111} (\bibinfo {year} {2019})}\BibitemShut {NoStop}%
\bibitem [{\citenamefont {Tomassetti}\ \emph {et~al.}(2024)\citenamefont
  {Tomassetti}, \citenamefont {Kafle}, \citenamefont {Marcial}, \citenamefont
  {Margine},\ and\ \citenamefont {Kolmogorov}}]{Tomassetti2024}%
  \BibitemOpen
  \bibfield  {author} {\bibinfo {author} {\bibfnamefont {C.~R.}\ \bibnamefont
  {Tomassetti}}, \bibinfo {author} {\bibfnamefont {G.~P.}\ \bibnamefont
  {Kafle}}, \bibinfo {author} {\bibfnamefont {E.~T.}\ \bibnamefont {Marcial}},
  \bibinfo {author} {\bibfnamefont {E.~R.}\ \bibnamefont {Margine}},\ and\
  \bibinfo {author} {\bibfnamefont {A.~N.}\ \bibnamefont {Kolmogorov}},\
  }\bibfield  {title} {\bibinfo {title} {Prospect of high-temperature
  superconductivity in layered metal borocarbides},\ }\href
  {https://doi.org/10.1039/D4TC00210E} {\bibfield  {journal} {\bibinfo
  {journal} {Journal of Materials Chemistry C}\ }\textbf {\bibinfo {volume}
  {12}},\ \bibinfo {pages} {4870} (\bibinfo {year} {2024})}\BibitemShut
  {NoStop}%
\bibitem [{\citenamefont {Verma}\ \emph {et~al.}(2003)\citenamefont {Verma},
  \citenamefont {Modak}, \citenamefont {Gaitonde}, \citenamefont {Rao},
  \citenamefont {Godwal},\ and\ \citenamefont {Gupta}}]{Verma2003}%
  \BibitemOpen
  \bibfield  {author} {\bibinfo {author} {\bibfnamefont {A.~K.}\ \bibnamefont
  {Verma}}, \bibinfo {author} {\bibfnamefont {P.}~\bibnamefont {Modak}},
  \bibinfo {author} {\bibfnamefont {D.~M.}\ \bibnamefont {Gaitonde}}, \bibinfo
  {author} {\bibfnamefont {R.~S.}\ \bibnamefont {Rao}}, \bibinfo {author}
  {\bibfnamefont {B.~K.}\ \bibnamefont {Godwal}},\ and\ \bibinfo {author}
  {\bibfnamefont {L.~C.}\ \bibnamefont {Gupta}},\ }\bibfield  {title} {\bibinfo
  {title} {{Possible high-temperature superconductivity in hole-doped
  MgB$_2$C$_2$}},\ }\href {https://doi.org/10.1209/epl/i2003-00592-1}
  {\bibfield  {journal} {\bibinfo  {journal} {Europhysics Letters (EPL)}\
  }\textbf {\bibinfo {volume} {63}},\ \bibinfo {pages} {743} (\bibinfo {year}
  {2003})}\BibitemShut {NoStop}%
\bibitem [{\citenamefont {Pham}\ and\ \citenamefont {Nguyen}(2023)}]{Pham2023}%
  \BibitemOpen
  \bibfield  {author} {\bibinfo {author} {\bibfnamefont {T.-T.}\ \bibnamefont
  {Pham}}\ and\ \bibinfo {author} {\bibfnamefont {D.-L.}\ \bibnamefont
  {Nguyen}},\ }\bibfield  {title} {\bibinfo {title} {{First-principles
  prediction of superconductivity in MgB$_3$C$_3$}},\ }\href
  {https://doi.org/10.1103/PhysRevB.107.134502} {\bibfield  {journal} {\bibinfo
   {journal} {Physical Review B}\ }\textbf {\bibinfo {volume} {107}},\ \bibinfo
  {pages} {134502} (\bibinfo {year} {2023})}\BibitemShut {NoStop}%
\bibitem [{\citenamefont {Span\`o}\ \emph {et~al.}(2005)\citenamefont
  {Span\`o}, \citenamefont {Bernasconi},\ and\ \citenamefont
  {Kopnin}}]{Spano2005}%
  \BibitemOpen
  \bibfield  {author} {\bibinfo {author} {\bibfnamefont {E.}~\bibnamefont
  {Span\`o}}, \bibinfo {author} {\bibfnamefont {M.}~\bibnamefont
  {Bernasconi}},\ and\ \bibinfo {author} {\bibfnamefont {E.}~\bibnamefont
  {Kopnin}},\ }\bibfield  {title} {\bibinfo {title} {{Electron-phonon
  interaction in hole-doped MgB$_2$C$_2$}},\ }\href
  {https://doi.org/10.1103/PhysRevB.72.014530} {\bibfield  {journal} {\bibinfo
  {journal} {Physical Review B}\ }\textbf {\bibinfo {volume} {72}},\ \bibinfo
  {pages} {014530} (\bibinfo {year} {2005})}\BibitemShut {NoStop}%
\bibitem [{\citenamefont {Mori}(2002)}]{Mori2002}%
  \BibitemOpen
  \bibfield  {author} {\bibinfo {author} {\bibfnamefont {T.}~\bibnamefont
  {Mori}},\ }\bibfield  {title} {\bibinfo {title} {{Investigation of
  Superconductivity in Isoelectronic and Related Compounds of MgB$_2$}},\
  }\href {https://doi.org/10.1143/JPSJS.71S.323} {\bibfield  {journal}
  {\bibinfo  {journal} {Journal of the Physical Society of Japan}\ }\textbf
  {\bibinfo {volume} {71}},\ \bibinfo {pages} {323} (\bibinfo {year}
  {2002})}\BibitemShut {NoStop}%
\bibitem [{\citenamefont {Mori}\ and\ \citenamefont
  {Takayama-Muromachi}(2004)}]{Mori2004}%
  \BibitemOpen
  \bibfield  {author} {\bibinfo {author} {\bibfnamefont {T.}~\bibnamefont
  {Mori}}\ and\ \bibinfo {author} {\bibfnamefont {E.}~\bibnamefont
  {Takayama-Muromachi}},\ }\bibfield  {title} {\bibinfo {title} {{Hole doping
  of MgB$_2$C$_2$, a MgB$_2$ related [B/C] layered compound}},\ }\href
  {https://doi.org/10.1016/j.cap.2003.11.027} {\bibfield  {journal} {\bibinfo
  {journal} {Current Applied Physics}\ }\textbf {\bibinfo {volume} {4}},\
  \bibinfo {pages} {276} (\bibinfo {year} {2004})}\BibitemShut {NoStop}%
\bibitem [{\citenamefont {Miao}\ \emph {et~al.}(2016)\citenamefont {Miao},
  \citenamefont {Huang},\ and\ \citenamefont {Yang}}]{Miao2016}%
  \BibitemOpen
  \bibfield  {author} {\bibinfo {author} {\bibfnamefont {R.}~\bibnamefont
  {Miao}}, \bibinfo {author} {\bibfnamefont {G.}~\bibnamefont {Huang}},\ and\
  \bibinfo {author} {\bibfnamefont {J.}~\bibnamefont {Yang}},\ }\bibfield
  {title} {\bibinfo {title} {{First-principles prediction of MgB$_2$-like NaBC:
  A more promising high-temperature superconducting material than LiBC}},\
  }\href {https://doi.org/10.1016/j.ssc.2016.02.011} {\bibfield  {journal}
  {\bibinfo  {journal} {Solid State Communications}\ }\textbf {\bibinfo
  {volume} {233}},\ \bibinfo {pages} {30} (\bibinfo {year} {2016})}\BibitemShut
  {NoStop}%
\bibitem [{\citenamefont {Singh}(2002)}]{Singh2002}%
  \BibitemOpen
  \bibfield  {author} {\bibinfo {author} {\bibfnamefont {P.~P.}\ \bibnamefont
  {Singh}},\ }\bibfield  {title} {\bibinfo {title} {{Hole-doped,
  high-temperature superconductors Li$_x$BC, Na$_x$BC and C$_x$: a
  coherent-potential-based prediction}},\ }\href
  {https://doi.org/10.1016/S0038-1098(02)00454-4} {\bibfield  {journal}
  {\bibinfo  {journal} {Solid State Communications}\ }\textbf {\bibinfo
  {volume} {124}},\ \bibinfo {pages} {25} (\bibinfo {year} {2002})}\BibitemShut
  {NoStop}%
\bibitem [{\citenamefont {Delacroix}\ \emph {et~al.}(2021)\citenamefont
  {Delacroix}, \citenamefont {Igoa}, \citenamefont {Song}, \citenamefont
  {Godec}, \citenamefont {Coelho-Diogo}, \citenamefont {Gervais}, \citenamefont
  {Rousse},\ and\ \citenamefont {Portehault}}]{Delacroix2021}%
  \BibitemOpen
  \bibfield  {author} {\bibinfo {author} {\bibfnamefont {S.}~\bibnamefont
  {Delacroix}}, \bibinfo {author} {\bibfnamefont {F.}~\bibnamefont {Igoa}},
  \bibinfo {author} {\bibfnamefont {Y.}~\bibnamefont {Song}}, \bibinfo {author}
  {\bibfnamefont {Y.~L.}\ \bibnamefont {Godec}}, \bibinfo {author}
  {\bibfnamefont {C.}~\bibnamefont {Coelho-Diogo}}, \bibinfo {author}
  {\bibfnamefont {C.}~\bibnamefont {Gervais}}, \bibinfo {author} {\bibfnamefont
  {G.}~\bibnamefont {Rousse}},\ and\ \bibinfo {author} {\bibfnamefont
  {D.}~\bibnamefont {Portehault}},\ }\bibfield  {title} {\bibinfo {title}
  {{Electron Precise Sodium Carbaboride Nanocrystals from Molten Salts: Single
  Sources to Boron Carbides}},\ }\href
  {https://doi.org/10.1021/acs.inorgchem.0c03501} {\bibfield  {journal}
  {\bibinfo  {journal} {Inorg. Chem.}\ }\textbf {\bibinfo {volume} {60}},\
  \bibinfo {pages} {4252} (\bibinfo {year} {2021})}\BibitemShut {NoStop}%
\bibitem [{\citenamefont {Kresse}\ and\ \citenamefont
  {Furthm{\"u}ller}(1996)}]{Kresse1996}%
  \BibitemOpen
  \bibfield  {author} {\bibinfo {author} {\bibfnamefont {G.}~\bibnamefont
  {Kresse}}\ and\ \bibinfo {author} {\bibfnamefont {J.}~\bibnamefont
  {Furthm{\"u}ller}},\ }\bibfield  {title} {\bibinfo {title} {{Efficient
  iterative schemes for $ab~initio$ total-energy calculations using a
  plane-wave basis set}},\ }\href {https://doi.org/10.1103/PhysRevB.54.11169}
  {\bibfield  {journal} {\bibinfo  {journal} {Phys. Rev. B}\ }\textbf {\bibinfo
  {volume} {54}},\ \bibinfo {pages} {11169} (\bibinfo {year}
  {1996})}\BibitemShut {NoStop}%
\bibitem [{\citenamefont {Bl{\"o}chl}(1994)}]{Blochl1994}%
  \BibitemOpen
  \bibfield  {author} {\bibinfo {author} {\bibfnamefont {P.~E.}\ \bibnamefont
  {Bl{\"o}chl}},\ }\bibfield  {title} {\bibinfo {title} {Projector
  augmented-wave method},\ }\href {https://doi.org/10.1103/PhysRevB.50.17953}
  {\bibfield  {journal} {\bibinfo  {journal} {Phys. Rev. B}\ }\textbf {\bibinfo
  {volume} {50}},\ \bibinfo {pages} {17953} (\bibinfo {year}
  {1994})}\BibitemShut {NoStop}%
\bibitem [{\citenamefont {Kolmogorov}\ and\ \citenamefont
  {Crespi}(2005)}]{ak06}%
  \BibitemOpen
  \bibfield  {author} {\bibinfo {author} {\bibfnamefont {A.~N.}\ \bibnamefont
  {Kolmogorov}}\ and\ \bibinfo {author} {\bibfnamefont {V.~H.}\ \bibnamefont
  {Crespi}},\ }\bibfield  {title} {\bibinfo {title} {Registry-dependent
  interlayer potential for graphitic systems},\ }\href
  {https://doi.org/10.1103/PhysRevB.71.235415} {\bibfield  {journal} {\bibinfo
  {journal} {Phys. Rev. B}\ }\textbf {\bibinfo {volume} {71}},\ \bibinfo
  {pages} {235415} (\bibinfo {year} {2005})}\BibitemShut {NoStop}%
\bibitem [{\citenamefont {Lebègue}\ \emph {et~al.}(2010)\citenamefont
  {Lebègue}, \citenamefont {Harl}, \citenamefont {Gould}, \citenamefont
  {Ángyán}, \citenamefont {Kresse},\ and\ \citenamefont
  {Dobson}}]{Lebegue2010}%
  \BibitemOpen
  \bibfield  {author} {\bibinfo {author} {\bibfnamefont {S.}~\bibnamefont
  {Lebègue}}, \bibinfo {author} {\bibfnamefont {J.}~\bibnamefont {Harl}},
  \bibinfo {author} {\bibfnamefont {T.}~\bibnamefont {Gould}}, \bibinfo
  {author} {\bibfnamefont {J.~G.}\ \bibnamefont {Ángyán}}, \bibinfo {author}
  {\bibfnamefont {G.}~\bibnamefont {Kresse}},\ and\ \bibinfo {author}
  {\bibfnamefont {J.~F.}\ \bibnamefont {Dobson}},\ }\bibfield  {title}
  {\bibinfo {title} {Cohesive properties and asymptotics of the dispersion
  interaction in graphite by the random phase approximation},\ }\href
  {https://doi.org/10.1103/PhysRevLett.105.196401} {\bibfield  {journal}
  {\bibinfo  {journal} {Phys. Rev. Lett.}\ }\textbf {\bibinfo {volume} {105}},\
  \bibinfo {pages} {196401} (\bibinfo {year} {2010})}\BibitemShut {NoStop}%
\bibitem [{\citenamefont {Ning}\ \emph {et~al.}(2022)\citenamefont {Ning},
  \citenamefont {Kothakonda}, \citenamefont {Furness}, \citenamefont {Kaplan},
  \citenamefont {Ehlert}, \citenamefont {Brandenburg}, \citenamefont {Perdew},\
  and\ \citenamefont {Sun}}]{rvv-1}%
  \BibitemOpen
  \bibfield  {author} {\bibinfo {author} {\bibfnamefont {J.}~\bibnamefont
  {Ning}}, \bibinfo {author} {\bibfnamefont {M.}~\bibnamefont {Kothakonda}},
  \bibinfo {author} {\bibfnamefont {J.~W.}\ \bibnamefont {Furness}}, \bibinfo
  {author} {\bibfnamefont {A.~D.}\ \bibnamefont {Kaplan}}, \bibinfo {author}
  {\bibfnamefont {S.}~\bibnamefont {Ehlert}}, \bibinfo {author} {\bibfnamefont
  {J.~G.}\ \bibnamefont {Brandenburg}}, \bibinfo {author} {\bibfnamefont
  {J.~P.}\ \bibnamefont {Perdew}},\ and\ \bibinfo {author} {\bibfnamefont
  {J.}~\bibnamefont {Sun}},\ }\bibfield  {title} {\bibinfo {title} {{Workhorse
  minimally empirical dispersion-corrected density functional with tests for
  weakly bound systems: r2SCAN+rVV10}},\ }\href
  {https://doi.org/10.1103/PhysRevB.106.075422} {\bibfield  {journal} {\bibinfo
   {journal} {Phys. Rev. B}\ }\textbf {\bibinfo {volume} {106}},\ \bibinfo
  {pages} {075422} (\bibinfo {year} {2022})}\BibitemShut {NoStop}%
\bibitem [{\citenamefont {Klime{\v{s}}}\ \emph {et~al.}(2011)\citenamefont
  {Klime{\v{s}}}, \citenamefont {Bowler},\ and\ \citenamefont
  {Michaelides}}]{optB86b}%
  \BibitemOpen
  \bibfield  {author} {\bibinfo {author} {\bibfnamefont {J.}~\bibnamefont
  {Klime{\v{s}}}}, \bibinfo {author} {\bibfnamefont {D.~R.}\ \bibnamefont
  {Bowler}},\ and\ \bibinfo {author} {\bibfnamefont {A.}~\bibnamefont
  {Michaelides}},\ }\bibfield  {title} {\bibinfo {title} {Van der {Waals}
  density functionals applied to solids},\ }\href
  {https://journals.aps.org/prb/abstract/10.1103/PhysRevB.83.195131} {\bibfield
   {journal} {\bibinfo  {journal} {Phys. Rev. B}\ }\textbf {\bibinfo {volume}
  {83}},\ \bibinfo {pages} {195131} (\bibinfo {year} {2011})}\BibitemShut
  {NoStop}%
\bibitem [{\citenamefont {Klimeš}\ \emph {et~al.}(2010)\citenamefont
  {Klimeš}, \citenamefont {Bowler},\ and\ \citenamefont
  {Michaelides}}]{optB88}%
  \BibitemOpen
  \bibfield  {author} {\bibinfo {author} {\bibfnamefont {J.}~\bibnamefont
  {Klimeš}}, \bibinfo {author} {\bibfnamefont {D.~R.}\ \bibnamefont
  {Bowler}},\ and\ \bibinfo {author} {\bibfnamefont {A.}~\bibnamefont
  {Michaelides}},\ }\bibfield  {title} {\bibinfo {title} {Chemical accuracy for
  the van der {Waals} density functional},\ }\href
  {https://doi.org/10.1088/0953-8984/22/2/022201} {\bibfield  {journal}
  {\bibinfo  {journal} {J. Phys. Condens. Matter}\ }\textbf {\bibinfo {volume}
  {22}},\ \bibinfo {pages} {022201} (\bibinfo {year} {2010})}\BibitemShut
  {NoStop}%
\bibitem [{\citenamefont {Furness}\ \emph {et~al.}(2020)\citenamefont
  {Furness}, \citenamefont {Kaplan}, \citenamefont {Ning}, \citenamefont
  {Perdew},\ and\ \citenamefont {Sun}}]{r2scan}%
  \BibitemOpen
  \bibfield  {author} {\bibinfo {author} {\bibfnamefont {J.~W.}\ \bibnamefont
  {Furness}}, \bibinfo {author} {\bibfnamefont {A.~D.}\ \bibnamefont {Kaplan}},
  \bibinfo {author} {\bibfnamefont {J.}~\bibnamefont {Ning}}, \bibinfo {author}
  {\bibfnamefont {J.~P.}\ \bibnamefont {Perdew}},\ and\ \bibinfo {author}
  {\bibfnamefont {J.}~\bibnamefont {Sun}},\ }\bibfield  {title} {\bibinfo
  {title} {{Accurate and Numerically Efficient r$^2$SCAN Meta-Generalized
  Gradient Approximation}},\ }\href
  {https://doi.org/10.1021/acs.jpclett.0c02405} {\bibfield  {journal} {\bibinfo
   {journal} {J. Phys. Chem. Lett.}\ }\textbf {\bibinfo {volume} {11}},\
  \bibinfo {pages} {8208} (\bibinfo {year} {2020})}\BibitemShut {NoStop}%
\bibitem [{\citenamefont {Monkhorst}\ and\ \citenamefont
  {Pack}(1976)}]{Monkhorst1976}%
  \BibitemOpen
  \bibfield  {author} {\bibinfo {author} {\bibfnamefont {H.~J.}\ \bibnamefont
  {Monkhorst}}\ and\ \bibinfo {author} {\bibfnamefont {J.~D.}\ \bibnamefont
  {Pack}},\ }\bibfield  {title} {\bibinfo {title} {Special points for
  {Brillouin}-zone integrations},\ }\href
  {https://journals.aps.org/prb/abstract/10.1103/PhysRevB.13.5188} {\bibfield
  {journal} {\bibinfo  {journal} {Phys. Rev. B}\ }\textbf {\bibinfo {volume}
  {13}},\ \bibinfo {pages} {5188} (\bibinfo {year} {1976})}\BibitemShut
  {NoStop}%
\bibitem [{\citenamefont {Hajinazar}\ \emph {et~al.}(2021)\citenamefont
  {Hajinazar}, \citenamefont {Thorn}, \citenamefont {Sandoval}, \citenamefont
  {Kharabadze},\ and\ \citenamefont {Kolmogorov}}]{maise}%
  \BibitemOpen
  \bibfield  {author} {\bibinfo {author} {\bibfnamefont {S.}~\bibnamefont
  {Hajinazar}}, \bibinfo {author} {\bibfnamefont {A.}~\bibnamefont {Thorn}},
  \bibinfo {author} {\bibfnamefont {E.~D.}\ \bibnamefont {Sandoval}}, \bibinfo
  {author} {\bibfnamefont {S.}~\bibnamefont {Kharabadze}},\ and\ \bibinfo
  {author} {\bibfnamefont {A.~N.}\ \bibnamefont {Kolmogorov}},\ }\bibfield
  {title} {\bibinfo {title} {{MAISE}: Construction of neural network
  interatomic models and evolutionary structure optimization},\ }\href
  {https://doi.org/https://doi.org/10.1016/j.cpc.2020.107679} {\bibfield
  {journal} {\bibinfo  {journal} {Comput. Phys. Commun.}\ }\textbf {\bibinfo
  {volume} {259}},\ \bibinfo {pages} {107679} (\bibinfo {year}
  {2021})}\BibitemShut {NoStop}%
\bibitem [{\citenamefont {Togo}\ and\ \citenamefont {Tanaka}(2015)}]{Togo2015}%
  \BibitemOpen
  \bibfield  {author} {\bibinfo {author} {\bibfnamefont {A.}~\bibnamefont
  {Togo}}\ and\ \bibinfo {author} {\bibfnamefont {I.}~\bibnamefont {Tanaka}},\
  }\bibfield  {title} {\bibinfo {title} {{First principles phonon calculations
  in materials science}},\ }\href
  {https://doi.org/10.1016/j.scriptamat.2015.07.021} {\bibfield  {journal}
  {\bibinfo  {journal} {Scr. Mater.}\ }\textbf {\bibinfo {volume} {108}},\
  \bibinfo {pages} {1} (\bibinfo {year} {2015})}\BibitemShut {NoStop}%
\bibitem [{\citenamefont {Kharabadze}\ \emph {et~al.}(2023)\citenamefont
  {Kharabadze}, \citenamefont {Meyers}, \citenamefont {Tomassetti},
  \citenamefont {Margine}, \citenamefont {Mazin},\ and\ \citenamefont
  {Kolmogorov}}]{Kharabadze2023}%
  \BibitemOpen
  \bibfield  {author} {\bibinfo {author} {\bibfnamefont {S.}~\bibnamefont
  {Kharabadze}}, \bibinfo {author} {\bibfnamefont {M.}~\bibnamefont {Meyers}},
  \bibinfo {author} {\bibfnamefont {C.~R.}\ \bibnamefont {Tomassetti}},
  \bibinfo {author} {\bibfnamefont {E.~R.}\ \bibnamefont {Margine}}, \bibinfo
  {author} {\bibfnamefont {I.~I.}\ \bibnamefont {Mazin}},\ and\ \bibinfo
  {author} {\bibfnamefont {A.~N.}\ \bibnamefont {Kolmogorov}},\ }\bibfield
  {title} {\bibinfo {title} {Thermodynamic stability of {Li–B–C} compounds
  from first principles},\ }\href {https://doi.org/10.1039/D2CP05500G}
  {\bibfield  {journal} {\bibinfo  {journal} {Phys. Chem. Chem. Phys.}\
  }\textbf {\bibinfo {volume} {25}},\ \bibinfo {pages} {7344} (\bibinfo {year}
  {2023})}\BibitemShut {NoStop}%
\bibitem [{\citenamefont {Thorn}\ \emph {et~al.}(2023)\citenamefont {Thorn},
  \citenamefont {Gochitashvili}, \citenamefont {Kharabadze},\ and\
  \citenamefont {Kolmogorov}}]{thorn2023}%
  \BibitemOpen
  \bibfield  {author} {\bibinfo {author} {\bibfnamefont {A.}~\bibnamefont
  {Thorn}}, \bibinfo {author} {\bibfnamefont {D.}~\bibnamefont
  {Gochitashvili}}, \bibinfo {author} {\bibfnamefont {S.}~\bibnamefont
  {Kharabadze}},\ and\ \bibinfo {author} {\bibfnamefont {A.~N.}\ \bibnamefont
  {Kolmogorov}},\ }\bibfield  {title} {\bibinfo {title} {{Machine learning
  search for stable binary Sn alloys with Na{,} Ca{,} Cu{,} Pd{,} and Ag}},\
  }\href {https://doi.org/10.1039/D3CP02817H} {\bibfield  {journal} {\bibinfo
  {journal} {Phys. Chem. Chem. Phys.}\ }\textbf {\bibinfo {volume} {25}},\
  \bibinfo {pages} {22415} (\bibinfo {year} {2023})}\BibitemShut {NoStop}%
\bibitem [{\citenamefont {Kolmogorov}\ \emph {et~al.}(2010)\citenamefont
  {Kolmogorov}, \citenamefont {Shah}, \citenamefont {Margine}, \citenamefont
  {Bialon}, \citenamefont {Hammerschmidt},\ and\ \citenamefont
  {Drautz}}]{ak16}%
  \BibitemOpen
  \bibfield  {author} {\bibinfo {author} {\bibfnamefont {A.~N.}\ \bibnamefont
  {Kolmogorov}}, \bibinfo {author} {\bibfnamefont {S.}~\bibnamefont {Shah}},
  \bibinfo {author} {\bibfnamefont {E.~R.}\ \bibnamefont {Margine}}, \bibinfo
  {author} {\bibfnamefont {A.~F.}\ \bibnamefont {Bialon}}, \bibinfo {author}
  {\bibfnamefont {T.}~\bibnamefont {Hammerschmidt}},\ and\ \bibinfo {author}
  {\bibfnamefont {R.}~\bibnamefont {Drautz}},\ }\bibfield  {title} {\bibinfo
  {title} {{New Superconducting and Semiconducting Fe-B Compounds Predicted
  with an Ab Initio Evolutionary Search}},\ }\href
  {https://doi.org/10.1103/PhysRevLett.105.217003} {\bibfield  {journal}
  {\bibinfo  {journal} {Phys. Rev. Lett.}\ }\textbf {\bibinfo {volume} {105}},\
  \bibinfo {pages} {217003} (\bibinfo {year} {2010})}\BibitemShut {NoStop}%
\bibitem [{\citenamefont {Giannozzi}\ \emph {et~al.}(2017)\citenamefont
  {Giannozzi}, \citenamefont {Andreussi}, \citenamefont {Brumme}, \citenamefont
  {Bunau}, \citenamefont {Nardelli}, \citenamefont {Calandra}, \citenamefont
  {Car}, \citenamefont {Cavazzoni}, \citenamefont {Ceresoli},\ and\
  \citenamefont {Cococcioni~$et. al$}}]{QE}%
  \BibitemOpen
  \bibfield  {author} {\bibinfo {author} {\bibfnamefont {P.}~\bibnamefont
  {Giannozzi}}, \bibinfo {author} {\bibfnamefont {O.}~\bibnamefont
  {Andreussi}}, \bibinfo {author} {\bibfnamefont {T.}~\bibnamefont {Brumme}},
  \bibinfo {author} {\bibfnamefont {O.}~\bibnamefont {Bunau}}, \bibinfo
  {author} {\bibfnamefont {M.~B.}\ \bibnamefont {Nardelli}}, \bibinfo {author}
  {\bibfnamefont {M.}~\bibnamefont {Calandra}}, \bibinfo {author}
  {\bibfnamefont {R.}~\bibnamefont {Car}}, \bibinfo {author} {\bibfnamefont
  {C.}~\bibnamefont {Cavazzoni}}, \bibinfo {author} {\bibfnamefont
  {D.}~\bibnamefont {Ceresoli}},\ and\ \bibinfo {author} {\bibfnamefont
  {M.}~\bibnamefont {Cococcioni~$et. al$}},\ }\bibfield  {title} {\bibinfo
  {title} {Advanced capabilities for materials modelling with {Quantum}
  {ESPRESSO}},\ }\href
  {https://iopscience.iop.org/article/10.1088/1361-648X/aa8f79/meta} {\bibfield
   {journal} {\bibinfo  {journal} {J. Phys: Condens. Matter}\ }\textbf
  {\bibinfo {volume} {29}},\ \bibinfo {pages} {465901} (\bibinfo {year}
  {2017})}\BibitemShut {NoStop}%
\bibitem [{\citenamefont {Thonhauser}\ \emph {et~al.}(2007)\citenamefont
  {Thonhauser}, \citenamefont {Cooper}, \citenamefont {Li}, \citenamefont
  {Puzder}, \citenamefont {Hyldgaard},\ and\ \citenamefont
  {Langreth}}]{Thonhauser2007}%
  \BibitemOpen
  \bibfield  {author} {\bibinfo {author} {\bibfnamefont {T.}~\bibnamefont
  {Thonhauser}}, \bibinfo {author} {\bibfnamefont {V.~R.}\ \bibnamefont
  {Cooper}}, \bibinfo {author} {\bibfnamefont {S.}~\bibnamefont {Li}}, \bibinfo
  {author} {\bibfnamefont {A.}~\bibnamefont {Puzder}}, \bibinfo {author}
  {\bibfnamefont {P.}~\bibnamefont {Hyldgaard}},\ and\ \bibinfo {author}
  {\bibfnamefont {D.~C.}\ \bibnamefont {Langreth}},\ }\bibfield  {title}
  {\bibinfo {title} {Van der {Waals} density functional: {Self}-consistent
  potential and the nature of the van der {Waals} bond},\ }\href
  {https://journals.aps.org/prb/abstract/10.1103/PhysRevB.76.125112} {\bibfield
   {journal} {\bibinfo  {journal} {Phys. Rev. B}\ }\textbf {\bibinfo {volume}
  {76}},\ \bibinfo {pages} {125112} (\bibinfo {year} {2007})}\BibitemShut
  {NoStop}%
\bibitem [{\citenamefont {Thonhauser}\ \emph {et~al.}(2015)\citenamefont
  {Thonhauser}, \citenamefont {Zuluaga}, \citenamefont {Arter}, \citenamefont
  {Berland}, \citenamefont {Schr{\"o}der},\ and\ \citenamefont
  {Hyldgaard}}]{Thonhauser2015}%
  \BibitemOpen
  \bibfield  {author} {\bibinfo {author} {\bibfnamefont {T.}~\bibnamefont
  {Thonhauser}}, \bibinfo {author} {\bibfnamefont {S.}~\bibnamefont {Zuluaga}},
  \bibinfo {author} {\bibfnamefont {C.~A.}\ \bibnamefont {Arter}}, \bibinfo
  {author} {\bibfnamefont {K.}~\bibnamefont {Berland}}, \bibinfo {author}
  {\bibfnamefont {E.}~\bibnamefont {Schr{\"o}der}},\ and\ \bibinfo {author}
  {\bibfnamefont {P.}~\bibnamefont {Hyldgaard}},\ }\bibfield  {title} {\bibinfo
  {title} {Spin signature of nonlocal correlation binding in metal-organic
  frameworks},\ }\href
  {https://journals.aps.org/prl/abstract/10.1103/PhysRevLett.115.136402}
  {\bibfield  {journal} {\bibinfo  {journal} {Phys. Rev. Lett.}\ }\textbf
  {\bibinfo {volume} {115}},\ \bibinfo {pages} {136402} (\bibinfo {year}
  {2015})}\BibitemShut {NoStop}%
\bibitem [{\citenamefont {Berland}\ \emph {et~al.}(2015)\citenamefont
  {Berland}, \citenamefont {Cooper}, \citenamefont {Lee}, \citenamefont
  {Schr{\"o}der}, \citenamefont {Thonhauser}, \citenamefont {Hyldgaard},\ and\
  \citenamefont {Lundqvist}}]{Berland2015}%
  \BibitemOpen
  \bibfield  {author} {\bibinfo {author} {\bibfnamefont {K.}~\bibnamefont
  {Berland}}, \bibinfo {author} {\bibfnamefont {V.~R.}\ \bibnamefont {Cooper}},
  \bibinfo {author} {\bibfnamefont {K.}~\bibnamefont {Lee}}, \bibinfo {author}
  {\bibfnamefont {E.}~\bibnamefont {Schr{\"o}der}}, \bibinfo {author}
  {\bibfnamefont {T.}~\bibnamefont {Thonhauser}}, \bibinfo {author}
  {\bibfnamefont {P.}~\bibnamefont {Hyldgaard}},\ and\ \bibinfo {author}
  {\bibfnamefont {B.~I.}\ \bibnamefont {Lundqvist}},\ }\bibfield  {title}
  {\bibinfo {title} {Van der {Waals} forces in density functional theory: a
  review of the {vdW-DF} method},\ }\href
  {https://iopscience.iop.org/article/10.1088/0034-4885/78/6/066501/meta}
  {\bibfield  {journal} {\bibinfo  {journal} {Rep. Prog. Phys.}\ }\textbf
  {\bibinfo {volume} {78}},\ \bibinfo {pages} {066501} (\bibinfo {year}
  {2015})}\BibitemShut {NoStop}%
\bibitem [{\citenamefont {Langreth}\ \emph {et~al.}(2009)\citenamefont
  {Langreth}, \citenamefont {Lundqvist}, \citenamefont {Chakarova-K{\"a}ck},
  \citenamefont {Cooper}, \citenamefont {Dion}, \citenamefont {Hyldgaard},
  \citenamefont {Kelkkanen}, \citenamefont {Kleis}, \citenamefont {Kong},\ and\
  \citenamefont {Li~$et. al$}}]{Langreth2009}%
  \BibitemOpen
  \bibfield  {author} {\bibinfo {author} {\bibfnamefont {D.~C.}\ \bibnamefont
  {Langreth}}, \bibinfo {author} {\bibfnamefont {B.~I.}\ \bibnamefont
  {Lundqvist}}, \bibinfo {author} {\bibfnamefont {S.~D.}\ \bibnamefont
  {Chakarova-K{\"a}ck}}, \bibinfo {author} {\bibfnamefont {V.~R.}\ \bibnamefont
  {Cooper}}, \bibinfo {author} {\bibfnamefont {M.}~\bibnamefont {Dion}},
  \bibinfo {author} {\bibfnamefont {P.}~\bibnamefont {Hyldgaard}}, \bibinfo
  {author} {\bibfnamefont {A.}~\bibnamefont {Kelkkanen}}, \bibinfo {author}
  {\bibfnamefont {J.}~\bibnamefont {Kleis}}, \bibinfo {author} {\bibfnamefont
  {L.}~\bibnamefont {Kong}},\ and\ \bibinfo {author} {\bibfnamefont
  {S.}~\bibnamefont {Li~$et. al$}},\ }\bibfield  {title} {\bibinfo {title} {A
  density functional for sparse matter},\ }\href
  {https://iopscience.iop.org/article/10.1088/0953-8984/21/8/084203/meta}
  {\bibfield  {journal} {\bibinfo  {journal} {J. Phys: Condens. Matter}\
  }\textbf {\bibinfo {volume} {21}},\ \bibinfo {pages} {084203} (\bibinfo
  {year} {2009})}\BibitemShut {NoStop}%
\bibitem [{\citenamefont {van Setten}\ \emph {et~al.}(2018)\citenamefont {van
  Setten}, \citenamefont {Giantomassi}, \citenamefont {Bousquet}, \citenamefont
  {Verstraete}, \citenamefont {Hamann}, \citenamefont {Gonze},\ and\
  \citenamefont {Rignanese}}]{Dojo2018}%
  \BibitemOpen
  \bibfield  {author} {\bibinfo {author} {\bibfnamefont {M.~J.}\ \bibnamefont
  {van Setten}}, \bibinfo {author} {\bibfnamefont {M.}~\bibnamefont
  {Giantomassi}}, \bibinfo {author} {\bibfnamefont {E.}~\bibnamefont
  {Bousquet}}, \bibinfo {author} {\bibfnamefont {M.~J.}\ \bibnamefont
  {Verstraete}}, \bibinfo {author} {\bibfnamefont {D.~R.}\ \bibnamefont
  {Hamann}}, \bibinfo {author} {\bibfnamefont {X.}~\bibnamefont {Gonze}},\ and\
  \bibinfo {author} {\bibfnamefont {G.-M.}\ \bibnamefont {Rignanese}},\
  }\bibfield  {title} {\bibinfo {title} {The {PseudoDojo}: {Training} and
  grading a 85 element optimized norm-conserving pseudopotential table},\
  }\href {https://doi.org/10.1016/j.cpc.2018.01.012} {\bibfield  {journal}
  {\bibinfo  {journal} {Comput. Phys. Commun.}\ }\textbf {\bibinfo {volume}
  {226}},\ \bibinfo {pages} {39} (\bibinfo {year} {2018})}\BibitemShut
  {NoStop}%
\bibitem [{\citenamefont {Perdew}\ \emph {et~al.}(1996)\citenamefont {Perdew},
  \citenamefont {Burke},\ and\ \citenamefont {Ernzerhof}}]{PBE}%
  \BibitemOpen
  \bibfield  {author} {\bibinfo {author} {\bibfnamefont {J.~P.}\ \bibnamefont
  {Perdew}}, \bibinfo {author} {\bibfnamefont {K.}~\bibnamefont {Burke}},\ and\
  \bibinfo {author} {\bibfnamefont {M.}~\bibnamefont {Ernzerhof}},\ }\bibfield
  {title} {\bibinfo {title} {Generalized gradient approximation made simple},\
  }\href {https://journals.aps.org/prl/abstract/10.1103/PhysRevLett.77.3865}
  {\bibfield  {journal} {\bibinfo  {journal} {Phys. Rev. Lett.}\ }\textbf
  {\bibinfo {volume} {77}},\ \bibinfo {pages} {3865} (\bibinfo {year}
  {1996})}\BibitemShut {NoStop}%
\bibitem [{\citenamefont {Methfessel}\ and\ \citenamefont
  {Paxton}(1989)}]{Methfessel1989}%
  \BibitemOpen
  \bibfield  {author} {\bibinfo {author} {\bibfnamefont {M.}~\bibnamefont
  {Methfessel}}\ and\ \bibinfo {author} {\bibfnamefont {A.~T.}\ \bibnamefont
  {Paxton}},\ }\bibfield  {title} {\bibinfo {title} {High-precision sampling
  for {Brillouin}-zone integration in metals},\ }\href
  {https://journals.aps.org/prb/abstract/10.1103/PhysRevB.40.3616} {\bibfield
  {journal} {\bibinfo  {journal} {Phys. Rev. B}\ }\textbf {\bibinfo {volume}
  {40}},\ \bibinfo {pages} {3616} (\bibinfo {year} {1989})}\BibitemShut
  {NoStop}%
\bibitem [{\citenamefont {Baroni}\ \emph {et~al.}(2001)\citenamefont {Baroni},
  \citenamefont {De~Gironcoli}, \citenamefont {Dal~Corso},\ and\ \citenamefont
  {Giannozzi}}]{Baroni2001}%
  \BibitemOpen
  \bibfield  {author} {\bibinfo {author} {\bibfnamefont {S.}~\bibnamefont
  {Baroni}}, \bibinfo {author} {\bibfnamefont {S.}~\bibnamefont
  {De~Gironcoli}}, \bibinfo {author} {\bibfnamefont {A.}~\bibnamefont
  {Dal~Corso}},\ and\ \bibinfo {author} {\bibfnamefont {P.}~\bibnamefont
  {Giannozzi}},\ }\bibfield  {title} {\bibinfo {title} {Phonons and related
  crystal properties from density-functional perturbation theory},\ }\href
  {https://journals.aps.org/rmp/abstract/10.1103/RevModPhys.73.515} {\bibfield
  {journal} {\bibinfo  {journal} {Rev. Mod. Phys.}\ }\textbf {\bibinfo {volume}
  {73}},\ \bibinfo {pages} {515} (\bibinfo {year} {2001})}\BibitemShut
  {NoStop}%
\bibitem [{SM()}]{SM}%
  \BibitemOpen
  \href@noop {} {\bibinfo  {journal} {{See Supplemental Material for
  Figs.~S1-S13 and Table~S1}}\ }\BibitemShut {NoStop}%
\bibitem [{\citenamefont {Giustino}\ \emph {et~al.}(2007)\citenamefont
  {Giustino}, \citenamefont {Cohen},\ and\ \citenamefont
  {Louie}}]{Giustino2007}%
  \BibitemOpen
\bibfield  {journal} {  }\bibfield  {author} {\bibinfo {author} {\bibfnamefont
  {F.}~\bibnamefont {Giustino}}, \bibinfo {author} {\bibfnamefont {M.~L.}\
  \bibnamefont {Cohen}},\ and\ \bibinfo {author} {\bibfnamefont {S.~G.}\
  \bibnamefont {Louie}},\ }\bibfield  {title} {\bibinfo {title}
  {Electron-phonon interaction using {Wannier} functions},\ }\href
  {https://journals.aps.org/prb/abstract/10.1103/PhysRevB.76.165108} {\bibfield
   {journal} {\bibinfo  {journal} {Phys. Rev. B}\ }\textbf {\bibinfo {volume}
  {76}},\ \bibinfo {pages} {165108} (\bibinfo {year} {2007})}\BibitemShut
  {NoStop}%
\bibitem [{\citenamefont {Ponc{\'e}}\ \emph {et~al.}(2016)\citenamefont
  {Ponc{\'e}}, \citenamefont {Margine}, \citenamefont {Verdi},\ and\
  \citenamefont {Giustino}}]{EPW2016}%
  \BibitemOpen
  \bibfield  {author} {\bibinfo {author} {\bibfnamefont {S.}~\bibnamefont
  {Ponc{\'e}}}, \bibinfo {author} {\bibfnamefont {E.~R.}\ \bibnamefont
  {Margine}}, \bibinfo {author} {\bibfnamefont {C.}~\bibnamefont {Verdi}},\
  and\ \bibinfo {author} {\bibfnamefont {F.}~\bibnamefont {Giustino}},\
  }\bibfield  {title} {\bibinfo {title} {{EPW}: {Electron}-phonon coupling,
  transport and superconducting properties using maximally localized {Wannier}
  functions},\ }\href
  {https://www.sciencedirect.com/science/article/pii/S0010465516302260}
  {\bibfield  {journal} {\bibinfo  {journal} {Comput. Phys. Commun.}\ }\textbf
  {\bibinfo {volume} {209}},\ \bibinfo {pages} {116} (\bibinfo {year}
  {2016})}\BibitemShut {NoStop}%
\bibitem [{\citenamefont {Margine}\ and\ \citenamefont
  {Giustino}(2013)}]{Margine2013}%
  \BibitemOpen
  \bibfield  {author} {\bibinfo {author} {\bibfnamefont {E.~R.}\ \bibnamefont
  {Margine}}\ and\ \bibinfo {author} {\bibfnamefont {F.}~\bibnamefont
  {Giustino}},\ }\bibfield  {title} {\bibinfo {title} {{Anisotropic
  Migdal-Eliashberg theory using Wannier functions}},\ }\href
  {https://journals.aps.org/prb/abstract/10.1103/PhysRevB.87.024505} {\bibfield
   {journal} {\bibinfo  {journal} {Phys. Rev. B}\ }\textbf {\bibinfo {volume}
  {87}},\ \bibinfo {pages} {024505} (\bibinfo {year} {2013})}\BibitemShut
  {NoStop}%
\bibitem [{\citenamefont {Lee}\ \emph {et~al.}(2023)\citenamefont {Lee},
  \citenamefont {Poncé}, \citenamefont {Bushick}, \citenamefont {Hajinazar},
  \citenamefont {Lafuente-Bartolome}, \citenamefont {Leveillee}, \citenamefont
  {Lian}, \citenamefont {Lihm}, \citenamefont {Macheda}, \citenamefont {Mori},
  \citenamefont {Paudyal}, \citenamefont {Sio}, \citenamefont {Tiwar},
  \citenamefont {Zacharias}, \citenamefont {Zhang}, \citenamefont {Bonini},
  \citenamefont {Kioupakis}, \citenamefont {Margine},\ and\ \citenamefont
  {Giustino}}]{EPW2023}%
  \BibitemOpen
  \bibfield  {author} {\bibinfo {author} {\bibfnamefont {H.}~\bibnamefont
  {Lee}}, \bibinfo {author} {\bibfnamefont {S.}~\bibnamefont {Poncé}},
  \bibinfo {author} {\bibfnamefont {K.}~\bibnamefont {Bushick}}, \bibinfo
  {author} {\bibfnamefont {S.}~\bibnamefont {Hajinazar}}, \bibinfo {author}
  {\bibfnamefont {J.}~\bibnamefont {Lafuente-Bartolome}}, \bibinfo {author}
  {\bibfnamefont {J.}~\bibnamefont {Leveillee}}, \bibinfo {author}
  {\bibfnamefont {C.}~\bibnamefont {Lian}}, \bibinfo {author} {\bibfnamefont
  {J.-M.}\ \bibnamefont {Lihm}}, \bibinfo {author} {\bibfnamefont
  {F.}~\bibnamefont {Macheda}}, \bibinfo {author} {\bibfnamefont
  {H.}~\bibnamefont {Mori}}, \bibinfo {author} {\bibfnamefont {H.}~\bibnamefont
  {Paudyal}}, \bibinfo {author} {\bibfnamefont {W.~H.}\ \bibnamefont {Sio}},
  \bibinfo {author} {\bibfnamefont {S.}~\bibnamefont {Tiwar}}, \bibinfo
  {author} {\bibfnamefont {M.}~\bibnamefont {Zacharias}}, \bibinfo {author}
  {\bibfnamefont {X.}~\bibnamefont {Zhang}}, \bibinfo {author} {\bibfnamefont
  {N.}~\bibnamefont {Bonini}}, \bibinfo {author} {\bibfnamefont
  {E.}~\bibnamefont {Kioupakis}}, \bibinfo {author} {\bibfnamefont {E.~R.}\
  \bibnamefont {Margine}},\ and\ \bibinfo {author} {\bibfnamefont
  {F.}~\bibnamefont {Giustino}},\ }\bibfield  {title} {\bibinfo {title}
  {{Electron-phonon physics from first principles using the EPW code}},\ }\href
  {https://doi.org/10.1038/s41524-023-01107-3} {\bibfield  {journal} {\bibinfo
  {journal} {npj Comput. Mater.}\ }\textbf {\bibinfo {volume} {9}},\ \bibinfo
  {pages} {156} (\bibinfo {year} {2023})}\BibitemShut {NoStop}%
\bibitem [{\citenamefont {Marzari}\ \emph {et~al.}(2012)\citenamefont
  {Marzari}, \citenamefont {Mostofi}, \citenamefont {Yates}, \citenamefont
  {Souza},\ and\ \citenamefont {Vanderbilt}}]{WANN1}%
  \BibitemOpen
  \bibfield  {author} {\bibinfo {author} {\bibfnamefont {N.}~\bibnamefont
  {Marzari}}, \bibinfo {author} {\bibfnamefont {A.~A.}\ \bibnamefont
  {Mostofi}}, \bibinfo {author} {\bibfnamefont {J.~R.}\ \bibnamefont {Yates}},
  \bibinfo {author} {\bibfnamefont {I.}~\bibnamefont {Souza}},\ and\ \bibinfo
  {author} {\bibfnamefont {D.}~\bibnamefont {Vanderbilt}},\ }\bibfield  {title}
  {\bibinfo {title} {Maximally localized {Wannier} functions: {Theory} and
  applications},\ }\href
  {https://journals.aps.org/rmp/abstract/10.1103/RevModPhys.84.1419} {\bibfield
   {journal} {\bibinfo  {journal} {Rev. Mod. Phys.}\ }\textbf {\bibinfo
  {volume} {84}},\ \bibinfo {pages} {1419} (\bibinfo {year}
  {2012})}\BibitemShut {NoStop}%
\bibitem [{\citenamefont {Pizzi}\ \emph {et~al.}(2020)\citenamefont {Pizzi},
  \citenamefont {Vitale}, \citenamefont {Arita}, \citenamefont {Bl{\"u}gel},
  \citenamefont {Freimuth}, \citenamefont {G{\'e}ranton}, \citenamefont
  {Gibertini}, \citenamefont {Gresch}, \citenamefont {Johnson},\ and\
  \citenamefont {Koretsune~$et. al$}}]{WANN2}%
  \BibitemOpen
  \bibfield  {author} {\bibinfo {author} {\bibfnamefont {G.}~\bibnamefont
  {Pizzi}}, \bibinfo {author} {\bibfnamefont {V.}~\bibnamefont {Vitale}},
  \bibinfo {author} {\bibfnamefont {R.}~\bibnamefont {Arita}}, \bibinfo
  {author} {\bibfnamefont {S.}~\bibnamefont {Bl{\"u}gel}}, \bibinfo {author}
  {\bibfnamefont {F.}~\bibnamefont {Freimuth}}, \bibinfo {author}
  {\bibfnamefont {G.}~\bibnamefont {G{\'e}ranton}}, \bibinfo {author}
  {\bibfnamefont {M.}~\bibnamefont {Gibertini}}, \bibinfo {author}
  {\bibfnamefont {D.}~\bibnamefont {Gresch}}, \bibinfo {author} {\bibfnamefont
  {C.}~\bibnamefont {Johnson}},\ and\ \bibinfo {author} {\bibfnamefont
  {T.}~\bibnamefont {Koretsune~$et. al$}},\ }\bibfield  {title} {\bibinfo
  {title} {Wannier90 as a community code: new features and applications},\
  }\href
  {https://iopscience.iop.org/article/10.1088/1361-648X/ab51ff?hootPostID=8865030f3411ebd77f127a8addfbbdce}
  {\bibfield  {journal} {\bibinfo  {journal} {J. Phys: Condens. Matter}\
  }\textbf {\bibinfo {volume} {32}},\ \bibinfo {pages} {165902} (\bibinfo
  {year} {2020})}\BibitemShut {NoStop}%
\bibitem [{\citenamefont {Marrazzo}\ \emph {et~al.}(2023)\citenamefont
  {Marrazzo}, \citenamefont {Beck}, \citenamefont {Margine}, \citenamefont
  {Marzari}, \citenamefont {Mostofi}, \citenamefont {Qiao}, \citenamefont
  {Souza}, \citenamefont {Tsirkin}, \citenamefont {Yates},\ and\ \citenamefont
  {Pizzi}}]{WANN3}%
  \BibitemOpen
  \bibfield  {author} {\bibinfo {author} {\bibfnamefont {A.}~\bibnamefont
  {Marrazzo}}, \bibinfo {author} {\bibfnamefont {S.}~\bibnamefont {Beck}},
  \bibinfo {author} {\bibfnamefont {E.~R.}\ \bibnamefont {Margine}}, \bibinfo
  {author} {\bibfnamefont {N.}~\bibnamefont {Marzari}}, \bibinfo {author}
  {\bibfnamefont {A.~A.}\ \bibnamefont {Mostofi}}, \bibinfo {author}
  {\bibfnamefont {J.}~\bibnamefont {Qiao}}, \bibinfo {author} {\bibfnamefont
  {I.}~\bibnamefont {Souza}}, \bibinfo {author} {\bibfnamefont {S.~S.}\
  \bibnamefont {Tsirkin}}, \bibinfo {author} {\bibfnamefont {J.~R.}\
  \bibnamefont {Yates}},\ and\ \bibinfo {author} {\bibfnamefont
  {G.}~\bibnamefont {Pizzi}},\ }\href {https://arxiv.org/abs/2312.10769}
  {\bibinfo {title} {{The Wannier-Functions Software Ecosystem for Materials
  Simulations}}} (\bibinfo {year} {2023}),\ \Eprint
  {https://arxiv.org/abs/2312.10769} {arXiv:2312.10769 [cond-mat.mtrl-sci]}
  \BibitemShut {NoStop}%
\bibitem [{\citenamefont {Lucrezi}\ \emph {et~al.}(2024)\citenamefont
  {Lucrezi}, \citenamefont {Ferreira}, \citenamefont {Hajinazar}, \citenamefont
  {Mori}, \citenamefont {Paudyal}, \citenamefont {Margine},\ and\ \citenamefont
  {Heil}}]{Lucrezi2024}%
  \BibitemOpen
  \bibfield  {author} {\bibinfo {author} {\bibfnamefont {R.}~\bibnamefont
  {Lucrezi}}, \bibinfo {author} {\bibfnamefont {P.~P.}\ \bibnamefont
  {Ferreira}}, \bibinfo {author} {\bibfnamefont {S.}~\bibnamefont {Hajinazar}},
  \bibinfo {author} {\bibfnamefont {H.}~\bibnamefont {Mori}}, \bibinfo {author}
  {\bibfnamefont {H.}~\bibnamefont {Paudyal}}, \bibinfo {author} {\bibfnamefont
  {E.~R.}\ \bibnamefont {Margine}},\ and\ \bibinfo {author} {\bibfnamefont
  {C.}~\bibnamefont {Heil}},\ }\bibfield  {title} {\bibinfo {title}
  {{Full-bandwidth anisotropic Migdal-Eliashberg theory and its application to
  superhydrides}},\ }\href {https://doi.org/10.1038/s42005-024-01528-6}
  {\bibfield  {journal} {\bibinfo  {journal} {Communications Physics}\ }\textbf
  {\bibinfo {volume} {7}},\ \bibinfo {pages} {33} (\bibinfo {year}
  {2024})}\BibitemShut {NoStop}%
\bibitem [{\citenamefont {Mori}\ \emph {et~al.}(2024)\citenamefont {Mori},
  \citenamefont {Nomoto}, \citenamefont {Arita},\ and\ \citenamefont
  {Margine}}]{mori2024}%
  \BibitemOpen
  \bibfield  {author} {\bibinfo {author} {\bibfnamefont {H.}~\bibnamefont
  {Mori}}, \bibinfo {author} {\bibfnamefont {T.}~\bibnamefont {Nomoto}},
  \bibinfo {author} {\bibfnamefont {R.}~\bibnamefont {Arita}},\ and\ \bibinfo
  {author} {\bibfnamefont {E.~R.}\ \bibnamefont {Margine}},\ }\href@noop {}
  {\bibinfo {title} {{Efficient anisotropic Migdal-Eliashberg calculations with
  the Intermediate Representation basis and Wannier interpolation}}} (\bibinfo
  {year} {2024}),\ \Eprint {https://arxiv.org/abs/2404.11528} {arXiv:2404.11528
  [cond-mat.supr-con]} \BibitemShut {NoStop}%
\bibitem [{\citenamefont {Kafle}\ \emph {et~al.}(2022)\citenamefont {Kafle},
  \citenamefont {Tomassetti}, \citenamefont {Mazin}, \citenamefont
  {Kolmogorov},\ and\ \citenamefont {Margine}}]{Kafle2022}%
  \BibitemOpen
  \bibfield  {author} {\bibinfo {author} {\bibfnamefont {G.~P.}\ \bibnamefont
  {Kafle}}, \bibinfo {author} {\bibfnamefont {C.~R.}\ \bibnamefont
  {Tomassetti}}, \bibinfo {author} {\bibfnamefont {I.~I.}\ \bibnamefont
  {Mazin}}, \bibinfo {author} {\bibfnamefont {A.~N.}\ \bibnamefont
  {Kolmogorov}},\ and\ \bibinfo {author} {\bibfnamefont {E.~R.}\ \bibnamefont
  {Margine}},\ }\bibfield  {title} {\bibinfo {title} {${Ab}~initio$ study of
  {Li-Mg-B} superconductors},\ }\href
  {https://doi.org/10.1103/PhysRevMaterials.6.084801} {\bibfield  {journal}
  {\bibinfo  {journal} {Phys. Rev. Mater.}\ }\textbf {\bibinfo {volume} {6}},\
  \bibinfo {pages} {084801} (\bibinfo {year} {2022})}\BibitemShut {NoStop}%
\bibitem [{\citenamefont {Momma}\ and\ \citenamefont {Izumi}(2011)}]{vesta}%
  \BibitemOpen
  \bibfield  {author} {\bibinfo {author} {\bibfnamefont {K.}~\bibnamefont
  {Momma}}\ and\ \bibinfo {author} {\bibfnamefont {F.}~\bibnamefont {Izumi}},\
  }\bibfield  {title} {\bibinfo {title} {{{\it VESTA3} for three-dimensional
  visualization of crystal, volumetric and morphology data}},\ }\href
  {https://doi.org/10.1107/S0021889811038970} {\bibfield  {journal} {\bibinfo
  {journal} {Journal of Applied Crystallography}\ }\textbf {\bibinfo {volume}
  {44}},\ \bibinfo {pages} {1272} (\bibinfo {year} {2011})}\BibitemShut
  {NoStop}%
\bibitem [{\citenamefont {Kawamura}(2019)}]{fermisurfer}%
  \BibitemOpen
  \bibfield  {author} {\bibinfo {author} {\bibfnamefont {M.}~\bibnamefont
  {Kawamura}},\ }\bibfield  {title} {\bibinfo {title} {{FermiSurfer:
  Fermi-surface viewer providing multiple representation schemes}},\ }\href
  {https://doi.org/https://doi.org/10.1016/j.cpc.2019.01.017} {\bibfield
  {journal} {\bibinfo  {journal} {Computer Physics Communications}\ }\textbf
  {\bibinfo {volume} {239}},\ \bibinfo {pages} {197} (\bibinfo {year}
  {2019})}\BibitemShut {NoStop}%
\bibitem [{\citenamefont {Ellgen}(2014)}]{chem-potential-textbook}%
  \BibitemOpen
  \bibfield  {author} {\bibinfo {author} {\bibfnamefont {P.~C.}\ \bibnamefont
  {Ellgen}},\ }\href
  {https://chem.libretexts.org/Bookshelves/Physical_and_Theoretical_Chemistry_Textbook_Maps/Book%3A_Thermodynamics_and_Chemical_Equilibrium_(Ellgen)/24%3A_Indistinguishable_Molecules_-_Statistical_Thermodynamics_of_Ideal_Gases/24.10%3A_The_Gibbs_Free_Energy_for_One_Mole_of_An_Ideal_Gas}
  {\bibinfo {title} {Thermodynamics and chemical equilibrium}} (\bibinfo {year}
  {2014})\BibitemShut {NoStop}%
\bibitem [{\citenamefont {Dumitric{\v a}}\ \emph {et~al.}(2002)\citenamefont
  {Dumitric{\v a}}, \citenamefont {Landis},\ and\ \citenamefont
  {Yakobson}}]{dumitrica2002}%
  \BibitemOpen
  \bibfield  {author} {\bibinfo {author} {\bibfnamefont {T.}~\bibnamefont
  {Dumitric{\v a}}}, \bibinfo {author} {\bibfnamefont {C.~M.}\ \bibnamefont
  {Landis}},\ and\ \bibinfo {author} {\bibfnamefont {B.~I.}\ \bibnamefont
  {Yakobson}},\ }\bibfield  {title} {\bibinfo {title} {Curvature-induced
  polarization in carbon nanoshells},\ }\href
  {https://doi.org/10.1016/S0009-2614(02)00820-5} {\bibfield  {journal}
  {\bibinfo  {journal} {Chemical Physics Letters}\ }\textbf {\bibinfo {volume}
  {360}},\ \bibinfo {pages} {182} (\bibinfo {year} {2002})}\BibitemShut
  {NoStop}%
\bibitem [{\citenamefont {Sun}\ \emph {et~al.}(2022)\citenamefont {Sun},
  \citenamefont {Zhang}, \citenamefont {Wang}, \citenamefont {Ho},
  \citenamefont {Mazin},\ and\ \citenamefont {Antropov}}]{descriptor-lambda}%
  \BibitemOpen
  \bibfield  {author} {\bibinfo {author} {\bibfnamefont {Y.}~\bibnamefont
  {Sun}}, \bibinfo {author} {\bibfnamefont {F.}~\bibnamefont {Zhang}}, \bibinfo
  {author} {\bibfnamefont {C.-Z.}\ \bibnamefont {Wang}}, \bibinfo {author}
  {\bibfnamefont {K.-M.}\ \bibnamefont {Ho}}, \bibinfo {author} {\bibfnamefont
  {I.~I.}\ \bibnamefont {Mazin}},\ and\ \bibinfo {author} {\bibfnamefont
  {V.}~\bibnamefont {Antropov}},\ }\bibfield  {title} {\bibinfo {title}
  {Electron-phonon coupling strength from ab initio frozen-phonon approach},\
  }\href {https://doi.org/10.1103/PhysRevMaterials.6.074801} {\bibfield
  {journal} {\bibinfo  {journal} {Phys. Rev. Mater.}\ }\textbf {\bibinfo
  {volume} {6}},\ \bibinfo {pages} {074801} (\bibinfo {year}
  {2022})}\BibitemShut {NoStop}%
\end{thebibliography}%

\end{document}